# Dispersal-based species pools as sources of connectivity area mismatches

Clémentine Préau[1], Nicolas Dubos[1], Maxime Lenormand[1], Pierre Denelle[1,2], Marine Le Louarn[1], Samuel Alleaume[1], Sandra Luque[1]

[1]INRAE, National Research Institute on Agriculture, Food & the Environment, TETIS Unit, Montpellier, France

[2] Biodiversity, Macroecology & Biogeography, University of Goettingen, Göttingen, Germany

**Abstract**

Context - Prioritising is likely to differ depending on the species considered for connectivity assessments, leading to a lack of consensual decisions for territorial planning.

Objectives - The objective was to assess the relevance of identifying priority areas for connectivity for groups of species based on common dispersal abilities. We aimed to assess the impact of target groups choices on predicted priority areas.

Method - The study was located at the Thau Lagoon territory to demonstrate the methodological approach. Ecological niche modelling was used to quantify species resistance and to identify suitable habitat patches. We coupled the least-cost path methodology with circuit theory to assess species connectivity. We classified connectivity from high to low levels and averaged the results by dispersal groups.

Results - We found important differences in identified priority areas between groups with dissimilar dispersal abilities, with little overlap between highly connected areas. We identified a gap between the level of protection of low dispersal species and highly connected areas. We found mismatches between existing corridors and connectivity in low dispersal species, and a greater impact in areas of expected urban sprawl projects on favourably connected areas for species with high dispersal capabilities.

Conclusion - We have demonstrated that a diversity of dispersal capacity ranges must be accounted for in order to identify ecological corridors in programmes that aim to restore habitat connectivity at territorial levels. Our findings are oriented to support the decisions of planning initiatives, at both local and regional scale.

**Graphical abstract**

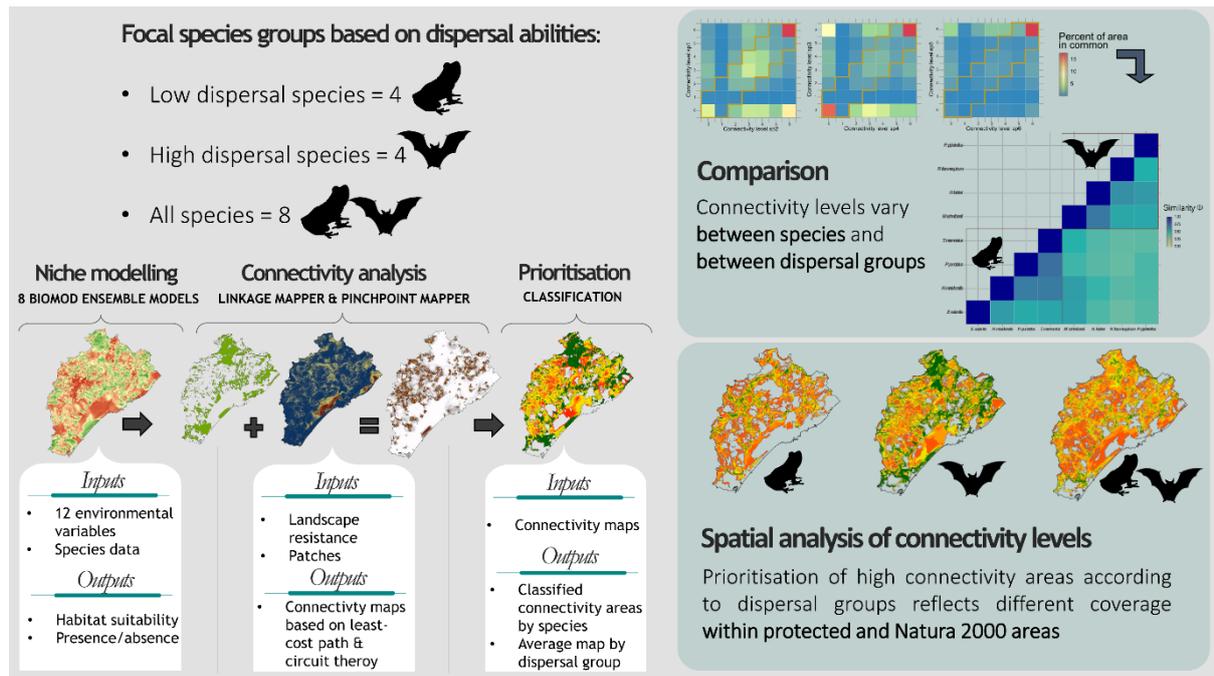

**Introduction**

Connectivity is a pressing concern for the majority of modern conservation initiatives around the globe, particularly in the light of the climate emergency and increasing land-use change (Luque et al. 2012; Keeley et al. 2018; Costanza and Terando 2019; Jalkanen et al. 2020). Connectivity, *i.e.,* the degree to which the landscape facilitates or impedes movement among vital patches (Taylor et al. 1993; Tischendorf and Fahrig 2000), is a critical issue when considering population viability, conditioning the movement of individuals, and observing gene flow between populations (*e.g.,* Stevens et al. 2006; Decout et al. 2012; Lizée et al. 2012; Albert et al. 2017; Le Roux et al. 2017). Habitat loss and fragmentation have therefore become major threats to biodiversity, mainly due to intensive agriculture, land encroachment (Kaim et al. 2019; WWF 2020), and landscape homogenisation (Allan et al. 2015; Buhk et al. 2017). Conservation plans promoting connectivity are notably oriented towards the implementation of ecological restoration actions, considering landscape structure, zoning areas for protection or management, and for guiding urban planning (Keeley et al. 2019). From a functional point of view, an effective conservation network should enable key life cycle phases, such as migration and dispersal, for multiple species that demonstrate various dispersal and habitat characteristics (Runge et al. 2014). Improving connectivity is therefore of fundamental importance to species resilience, given the current context of land use and land cover change, and potential range shifts induced by climate change (Rudnick et al. 2012; Costanza and Terando 2019). The prioritised conservation of areas that improve landscape connectivity has therefore become integral to the wider conservation agenda, and more broadly to the study of landscapes.

The establishment of a conservation network cannot explicitly account for the connectivity needs of every species. Network identification must therefore involve choices and compromises towards the implementation of operational methodological approaches which allow for the relevant species to be considered in the planning process. For instance, network identification can be achieved through the linkage of land units based on the continuity and composition of landscape structures. Whilst this method is commonly used in urban planning, it does not however directly account for species movement, such as dispersal and migration characteristics (Blasi et al. 2008; Biondi et al. 2012; Runge et al. 2014). A more holistic approach instead optimizes linkages between environmentally similar habitats after the identification of barriers, in order to determine corridors that are potentially suitable to numerous species (Alagador et al. 2012). Another common approach utilises the selection of a number of umbrella species such that the "conservation or restoration of dispersal habitat also facilitates dispersal of other target species" (Breckheimer et al. 2014). Many studies have assessed connectivity for a single charismatic species, such as large-bodied species or long-distance dispersers, in order to identify conservation corridors (Correa Ayram et al. 2016). Enhancing connectivity for co-occurring species however is challenging, as landscape resistance (*i.e.,* the cost of moving through the environment) greatly differs across species, and is dependent on species characteristics, habitat use, or dispersal abilities (Zeller et al. 2012; Richardson 2012). The capacity of a species to be a suitable surrogate not only depends on species characteristics but also on landscape features such as the amount of habitat and the level of fragmentation, which differ according to sites and dynamics over time (Diniz et al. 2018). Nevertheless, numerous authors have investigated multispecies approaches that overcome these limitations. The focal species approach targets taxa for the management of threatening processes (Lambeck 1997). The ability of species to perform as surrogates when they are either area-limited, dispersal-limited, resource-limited, or ecological process-limited has proven to be incomplete (Lindenmayer et al. 2002; Dondina et al. 2020). Many authors now advocate the identification of corridors using surrogate species that denote various taxa, namely a variety of habitat use, movement, or dispersal abilities, or a combination of the three (*e.g.,* Meurant et al. 2018).

Several approaches are commonly employed to identify ecological corridors such as least-cost analysis (Adriaensen et al. 2003), resistant kernel, individual-based dispersal models (Diniz et al. 2020), graph theory (Urban and Keitt 2001; Brás et al. 2013) or circuit theory (McRae et al. 2008). Circuit theory is a powerful conservation tool which considers several possible pathways through the landscape using a random walk modelling approach (McRae et al. 2008; Dickson et al. 2019). Although it varies widely between taxa, dispersal is a key element in connectivity assessment, regarding many parameters such as morphology, life cycle, mode of displacement, and perception of the landscape or landscape permeability, resulting in diverse displacement patterns (Baguette et al. 2013; Diniz et al. 2020). Hence, depending on the species considered for connectivity assessment, the prioritisation of conservation areas is likely to differ, leading to a lack of consensual decisions for territorial planning.

Here, we aimed to compare the relevance of identifying priority areas for connectivity based on groups of species selected according to similar dispersal abilities against prioritisation based on species having both high and low dispersal abilities. To this aim, we used ecological niche modelling to quantify how a landscape impedes movement by a species and to identify core areas according to habitat suitability. We then assessed species connectivity through least-cost path coupled with circuit theory for each species. Finally, we used a classification method to identify levels of connectivity and averaged the results by group, before to compare them on existing network of protected and Natura 2000 areas. The objective of the study was oriented to assess the impact of species target group selection on predicted priority areas for conservation. The approach provides a testing ground based on a French case as a demonstrator oriented to support decisions in territorial planning initiatives to balance conservation and urban development pressures.

**Material and methods**

**Study area**

The study was conducted at the "Thau Lagoon" territory located in the south of France, in the region Occitanie, on the Mediterranean coast (Fig. 1). The area is approximately 71,750 ha and provides a great summary of the diversity of Mediterranean landscapes, coastal plain, wooded reliefs, lagoons, wetlands, and agricultural plains. The area is subjected to several protection and conservation measures, such as the classification of large part of the territory as Natura 2000 sites (European Commission), a National Nature Reserve in the southwest and an area of biotope protection in the southeast. Nevertheless, this attractive area is subject to important anthropogenic pressures, including urban sprawl. This territory is regulated by an urban planning document, known as Territorial Coherence Plan (SCOT, Schéma de COhérence Territoriale), which defines a reference framework organising sectoral policies, in particular on spatial organisation and urban planning, including biodiversity, energy and climate (Ministère de la Cohésion des territoires et des Relations avec les collectivités territoriales).

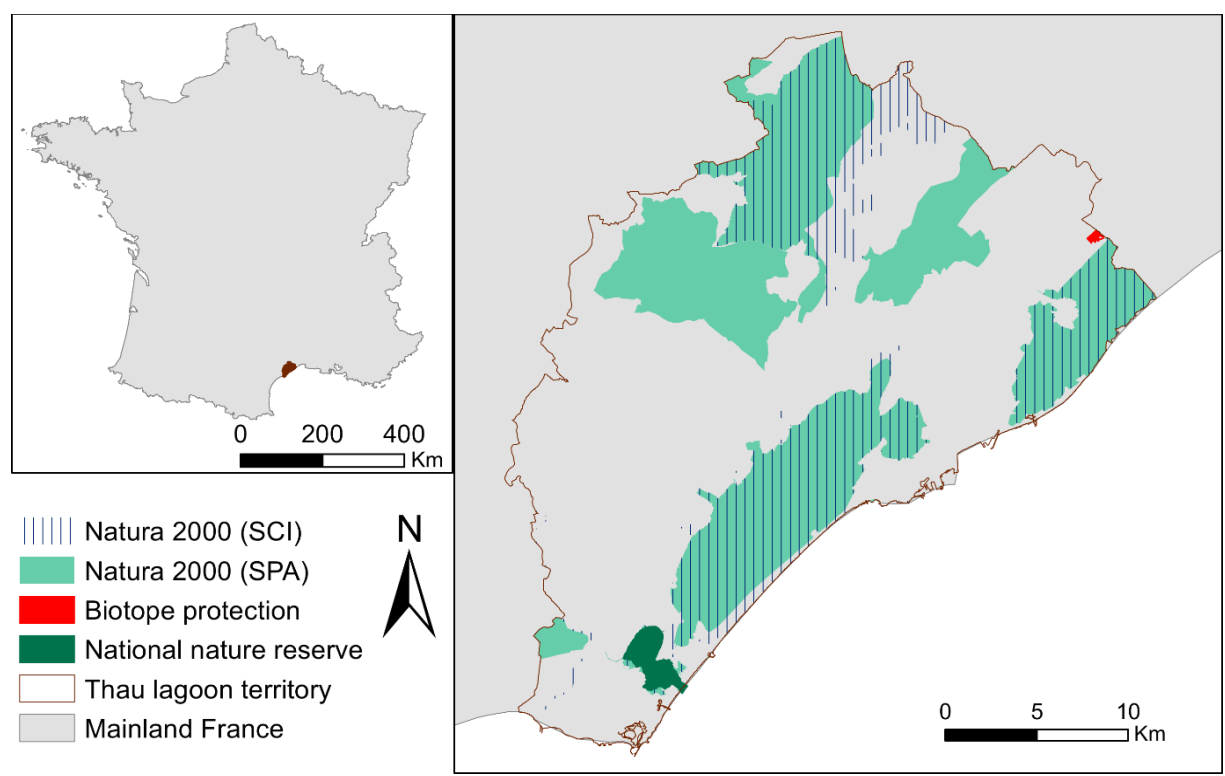

**Fig. 1** Map of the study area. SCI stands for Sites of Community Importance (Natura 2000), SPA stands for Special Protection Areas (Natura 2000)

**Study species and dispersal groups**

We selected groups of species based on dispersal abilities. We chose to focus on two taxonomic groups with very distinct environmental niches and extremely different movement capacities and needs. Within that rationality, we selected a group of amphibian species as a low dispersal group, a group of bat species as high dispersal group, then we combined them as a mixed group composed of all species. Four species of amphibians, *Triturus*

*marmoratus*, *Hyla meridionalis*, *Epidalea calamita* and *Pelodytes punctatus* were considered for a low dispersal group, and *Rhinolophus ferrumequinum*, *Miniopterus schreibersii*, *Nyctalus leisleri* and *Pipistrellus pipistrellus*, for a high dispersal group. While estimating the dispersal distances of species is challenging and may lead to underestimates (Santini et al. 2013; Fonte et al. 2019; Tournayre et al. 2019), we deliberately performed our analyses while taking into account long distances for both groups. The aim was to consider the different life cycle processes of the species such as dispersal and migration (Runge et al. 2014). We estimated ranges of species movement abilities from published studies (see Online Resource 1 for species information and references). Species occurrences, distributed from east to west over the study area, were compiled from several existing databases of two local NGOs: The French League for the Protection of Birds (LPO Occitanie) and the Chiroptera group of Languedoc Roussillon (GCLR). Bat occurrence data did not encompass roosts sites, therefore represented only activity periods (Le Roux et al. 2017).

**Ecological niche modelling**

We modelled the potential distribution of the eight species using Biomod 2 (Thuiller et al. 2009) under R software version 3.6.0 (R Core Team 2020), at a spatial resolution of 25x25m resulting in 2,242,338 grid cells. We ran the models with 12 environmental variables: distance to urban areas, density of built-up areas, elevation, distance to water bodies, distance to water courses, distance to intensive agriculture, distance to non-intensive agriculture, distance to coniferous forests, distance to deciduous forests, distance to mixed forests, distance to scrublands and annual maximum of Normalized Difference Vegetation Index (NDVI) (see Online Resource 2 for details about variables). We used Variance Inflation Factor (VIF) to assess collinearity among the predictors (we assumed a set of uncorrelated predictors when VIF<10, as recommended by Chatterjee and Hadi (2006); Naimi et al. (2014), and Naimi and Araújo (2016)). The actual VIF values for the 12 predictors varied from 1.291 to 3.225. To account for the spatial bias often inherent to naturalist databases (Dubos et al 2021), we created a dataset of pseudo-absences (PsA) for each species, with the same bias as occurrence data (Phillips et al. 2009) using probabilities based on a kernel density surface displaying higher probabilities in areas with more occurrence data (Fitzpatrick et al. 2013). We selected the same number of PsA than occurrence data. We first computed the models with the algorithms available in the package Biomod 2: GLM, GBM, GAM, CTA, SRE, ANN, FDA, RF, MARS and MAXENT.Phillips, with default options and all 12 predictor variables. For each species, we selected the eight most important predictor variables and reran the models with the selected predictors, following the recommendation of Brun et al. (2020) to use a reasonably small number of predictors. We used 50 runs per model, applying equal weight for presence and PsA. We used 70 % of datasets for training and 30 % for evaluation. We computed ensemble modelling though the mean of all models with True Skills Statistics (TSS) superior to 0.6. We evaluated the ensemble models with AUC-ROC, TSS, Kappa and Boyce Index (Hanley and McNeil 1982; Boyce et al. 2002; Hirzel et al. 2006; Allouche et al. 2006). We forecast the ensemble models to obtain a continuous prediction raster of habitat suitability (HS, ranging from 0 to 1), displaying the potential distribution of each species.

**Connectivity analysis**

We transformed the continuous predictions of HS into binary predictions according to the $10^{th}$ percentile threshold. This threshold allows to exclude areas with values of HS lower than the lowest 10 % of occurrence data, assuming that those localities are not representative of the habitat of the species. It has been widely adopted and has shown to perform well in identifying habitat patches (Pearson et al. 2007; Capinha et al. 2013; Ramirez-Reyes et al. 2016). For each species, we defined core areas as regions with HS superior to the threshold. We excluded core areas under a minimum size of 1 ha of the analysis. Then, for each species, we used the continuous predictions of HS to compute a resistance surface ranging from 1 to 1,000 through a negative exponential function, so that resistance=1 when HS was superior or equal to the 10th percentile threshold, and resistance=$e^{((\ln(0.001)/threshold) \times HS)} \times 10^3$ when HS was inferior to the threshold (Duflot et al. 2018). We automatically set urban areas and roads as high resistance pixels (*i.e.*, 1,000).

We conducted connectivity analysis using Linkage Mapper toolkit 2.0.0 (McRae and Kavanagh 2011) under ArcGIS 10.8 to compute least-cost paths between core areas. We built networks and mapped links using cost-weighted distances based on the resistance surfaces. We restricted link length to the approximated maximum dispersal distances for each species or species group according to the scenario (Online Resource 1). Then we ran PinchPoint Mapper tool, which calls Circuitscape to identify pinch-points (*i.e.*, bottlenecks) within least-cost corridors previously identified with Linkage Mapper, using circuit theory (McRae et al. 2008; McRae and Kavanagh 2011). This method enables maps of current flow to be produced, which identify areas where corridors are constricted and could be considered conservation priorities as small loss of habitat could greatly compromise connectivity. We used a large cut-off of 25 km in cost-weighted distance for corridor width. This would allow areas of relatively uniform width between species to be obtained, despite differences in resistance area values, as well as enabling the identification of corridors that may also be suitable for additional species other than the eight studied here. Furthermore, this made it possible to account for some inaccuracies, particularly those related to certain parameters from GIS data inputs (WHCWG 2010).

**Classification of connectivity levels**

For each species, we reclassified the outputs from PinchPoint Mapper in six levels of connectivity, where level 0 is equal to habitat patches and levels 1, 2, 3, 4 and 5 stand for corridor levels, from high to low connectivity potential (values of 6 and 7 respectively represented areas with no connectivity or outside the calculation area of Pinchpoint mapper). We assigned the thresholds for defining levels using a non-parametric method based on the derivative of the Lorenz curve (Louail et al. 2014; Bassolas et al. 2019)(Online Resource 3). We built a matrix H with element $H_{ij}$ corresponding to the number of shared pixels values between the resulting levels of connectivity assigned for each species. We then computed a metric of similarity Φ (ranging from 0 to 1) between species maps of connectivity levels, defined as the tri-diagonal trace of H, *i.e.,* the main diagonal, the first diagonal below this, and the first diagonal above the main diagonal only (see graphical explanation in Online Resource 4):

$$\Phi = \sum_{i,j=0}^{6} H_{ij}(\delta_{ij} + \delta_{i(j-1)} + \delta_{(i-1)j})$$

where $\delta_{ij}$ is the Kronecker delta (Bassolas et al. 2019). The similarity metric is useful in this application as it considers similarity within the same connectivity class, along with those connectivity classes immediately above and below it. We converted the pairwise matrices of shared pixel values to the percentage of surface area in common to simplify their interpretation (Online Resource 5).

We investigated the effect of species choice on the identification of target areas for conservation using three groups, namely: low dispersal species, high dispersal species and all species pooled together. For the three groups, we generated maps according to connectivity levels by averaging 1) connectivity scores for the four low dispersal species, 2) the four high dispersal species, 3) all eight species. We assessed confidence in averaged maps using standard deviation maps for the three groups. We reclassified averaged maps from high connectivity areas (0 to 1) to low connectivity areas (4 to 5). Within the limits of the SCOT. At the end, we identified the amount of area covered by protected and Natura 2000 areas; considering different land use types; according to existing corridors; and by areas identified for urban sprawl projects.

# Results

**Connectivity for dispersal groups**

Assemblages of ecological niche models resulted in satisfying evaluations by AUC (ranging from 0.943 for *P. pipistrellus* to 0.996 for *N. leisleri*), TSS (ranging from 0.753 for *P. pipistrellus* to 0.953 for *T. marmoratus*), and Boyce Index (ranging from 0.76 for *T. marmoratus* to 0.978 for *P. punctatus*) (Online Resource 1). After computing connectivity analysis based on the outputs of niche modelling, we averaged maps by groups and identified connectivity levels.

We consistently identified areas with an important potential for connectivity based on single groups. Though, we found more variability when accounting for multiple dispersal abilities (Fig. 2). Indeed, we found an important overall similarity in the classification of connectivity levels between the two groups, ranging from 0.52 to 0.7 between maps for amphibians (low dispersal group) and from 0.53 to 0.67 between maps for bats (high dispersal group). Similarity was smaller between species from the different groups, where it ranged from 0.4 to 0.56 (Fig. 2). With pairwise species assessment, we found the maximum percentage of shared connectivity values in the no connectivity areas (6) (from 33.2 % to 51.42 % for amphibians, from 17.5 % to 22.65 % for bats) (Online Resource 5). We found slight overlap between levels, especially between amphibian species and between amphibians and bats (Online Resource 5). The overlap was somewhat greater between bat species, in particular for level 0 corresponding to patches, in particular between *P. pipistrellus* and *N. leisleri* and between *P. pipistrellus* and *R. ferrumequinum*.

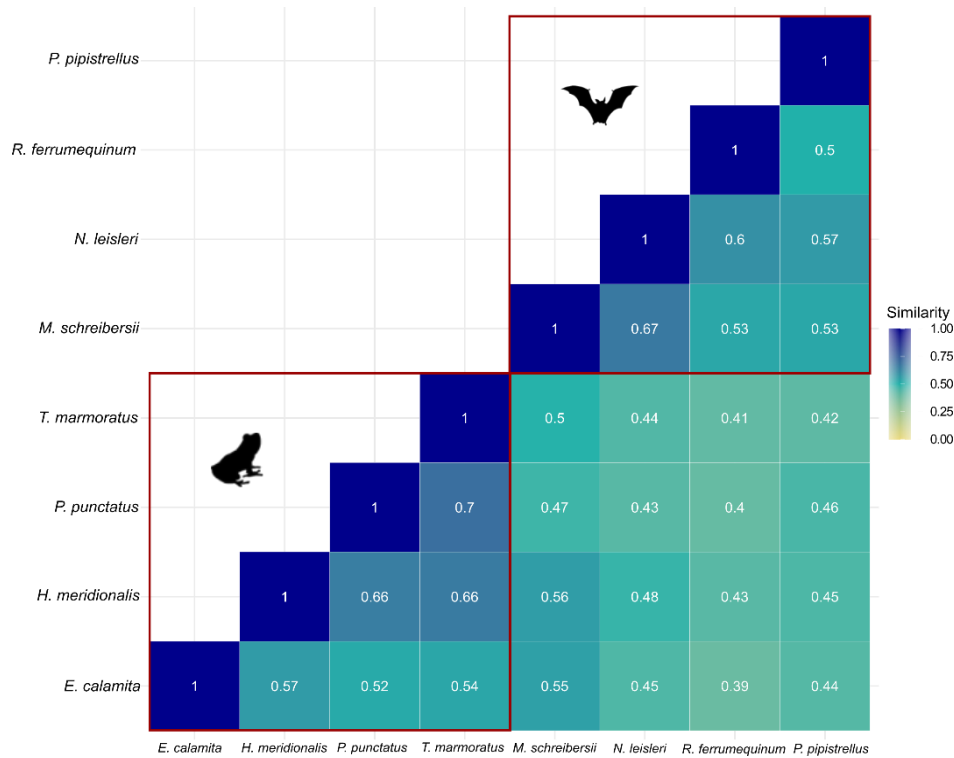

**Fig. 2** Similarity of connectivity levels among all species

We found similar patterns in connectivity levels in low dispersal and within high dispersal species, while the two groups showed different patterns (Online Resource 6). For amphibians, we identified unconnected large areas (from 47.69 % for *E. calamita* to 69.1 % for *T. marmoratus*), whereas connectivity levels covered a larger part of the study area for bat species. In particular, high connectivity levels, *i.e.*, patches, covered large areas for three of the four bat species (36.75 %, 30.85 % and 21.51 % of the study area for *P. pipistrellus*, *R. ferrumequinum* and *N. leisleri* respectively), in contrast to those of amphibians which ranged from 15.89 % of the study area for *E. calamita* to 0.81 % for *T. marmoratus*. We found different connectivity patterns between average maps of the two dispersal groups (low and high) (Fig. 3). Connectivity levels averaged by group allowed the identification of consistent areas between species of same low, high and both dispersal groups. Areas with important connectivity levels covered more surface for the high dispersal group than for the low dispersal group and all species group (12.56 % versus 0.75 % for low dispersal and 0.72 % for all species). Connectivity areas of the high dispersal group covered only 22.4 % of those identified for the low dispersal group against 93 % of those identified for all species.

Overall, the confidence (*i.e.,* inverse of standard deviation) between average maps was greater within groups of low and high dispersal than between all species. We identified the most isolated areas (*i.e.,* no connectivity) with a greater level of confidence (Online Resource 7).

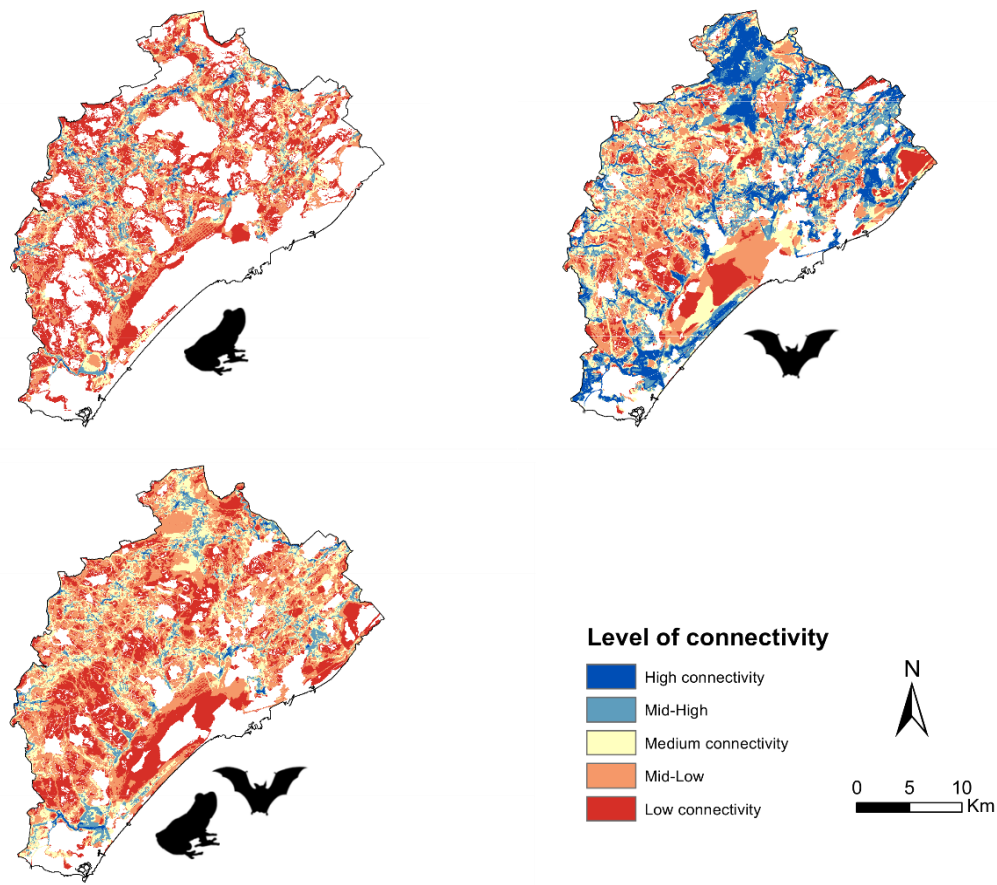

**Fig. 3** Average maps for the three groups. White areas correspond to no connectivity level

**Gap analysis**

The National Nature Reserve encompassed an important connectivity area for all three groups, especially for the high dispersal group. The area of biotope protection was more suitable for connectivity in the high dispersal than the low dispersal group. While Natura 2000 areas, covered a large part of the study area and did not seem to overlap highly connected areas better than the whole study region for either of the dispersal groups. Connectivity in Natura 2000 areas was assessed for both Sites of Community Importance (1992 Habitats Directive) and for Special Protection Areas (1979 Birds Directive) (Fig. 4). For both types of Natura 2000 areas, we globally found small levels to no connectivity for the low dispersal group, average connectivity for the high dispersal group and mid-low connectivity when accounting for all species in the mixed dispersal group.

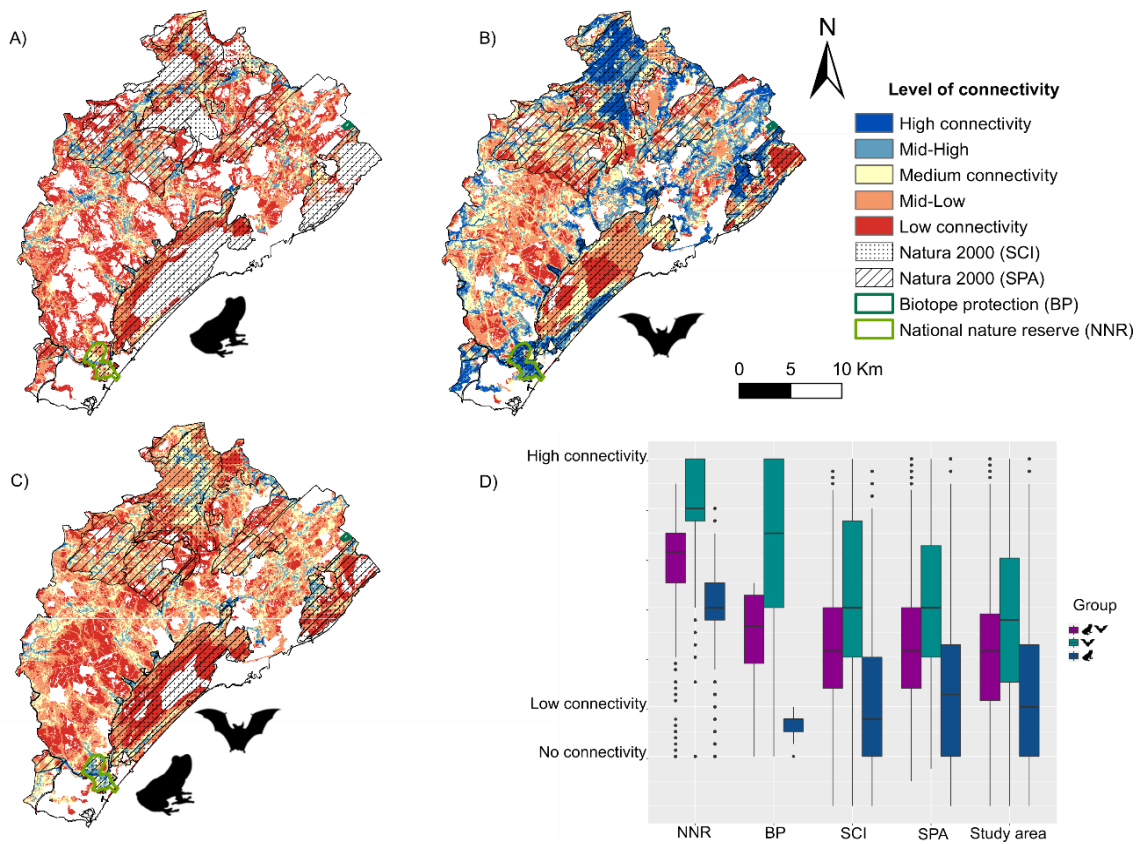

**Fig. 4** Levels of connectivity and covering by protected and Natura 2000 areas for low dispersal (A), high dispersal (B) and all species (C) groups within the study area; connectivity values within protected and Natura 2000 areas, and within the study area for the three groups (D). NNR stands for National Nature Reserve, BP stands for Biotope Protection, SCI stands for Sites of Community Importance (Natura 2000), SPA stands for Special Protection Areas (Natura 2000). White areas correspond to no connectivity level

The analysis of the most represented land uses according to the levels of connectivity identified in the study showed that the most favourable areas for connectivity correspond mainly to the same land use classes, independently of the dispersal capacity (Fig. 5). For each group, the land use classes most represented in these best-connected areas (dark and light blue on Fig. 5) were scrubland areas, permanent crops, forests, and maritime wetlands. On the contrary, the land-use classes corresponding to artificial territories mostly corresponded to areas of low connectivity levels. The corridors identified in the SCOT planning scheme for the Thau territory, showed different levels of connectivity according to species groups. They appear to be more suitable to the connectivity of the high dispersal group than to the low dispersal group, for which the main part of the area identified as corridors is classified as medium to low connectivity. Similarly, when all species are considered, a limited portion of the corridors are concerned with high connectivity levels (Fig. 6).

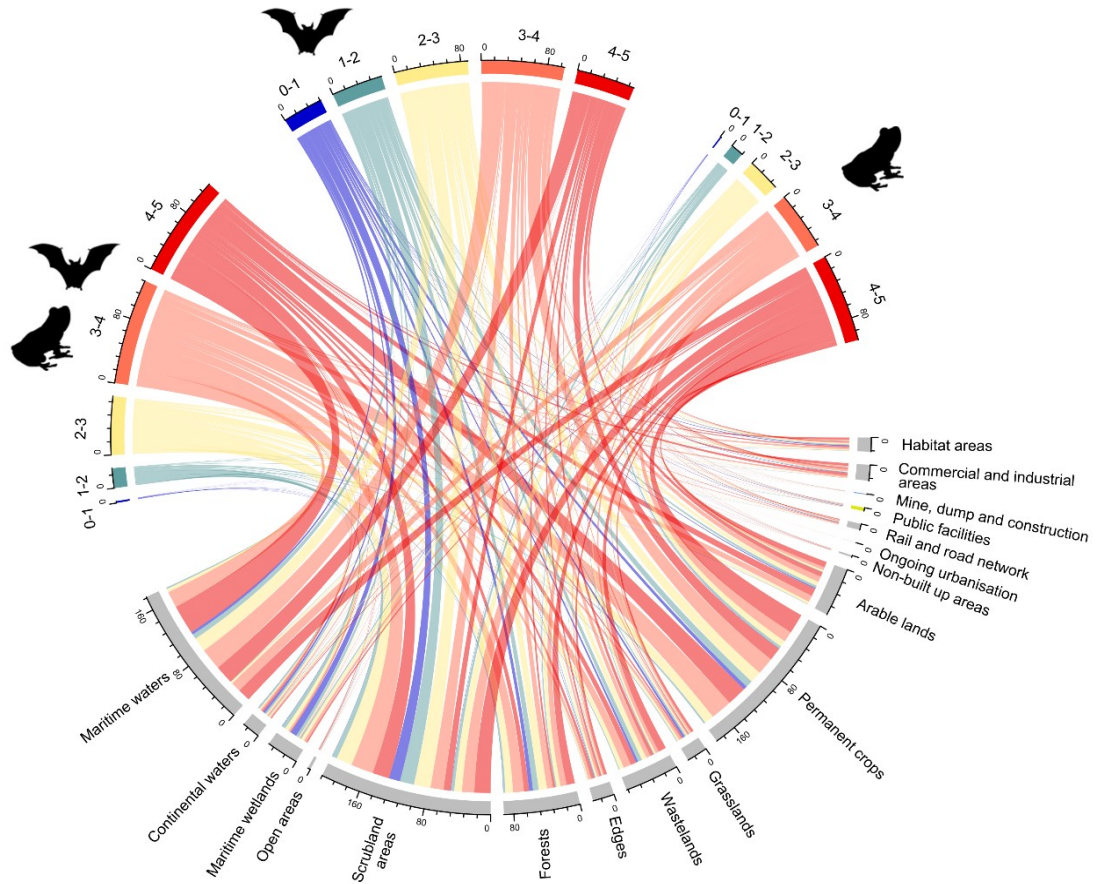

**Fig. 5** Allocation of surface areas of land use types covered by the different levels of connectivity identified for the three groups. Blue corresponds to high connectivity (0-1), light blue corresponds to mid-high connectivity (1-2), yellow corresponds to medium connectivity (2-3), orange corresponds to mid-low connectivity (3-4) and red corresponds to low connectivity (4-5)

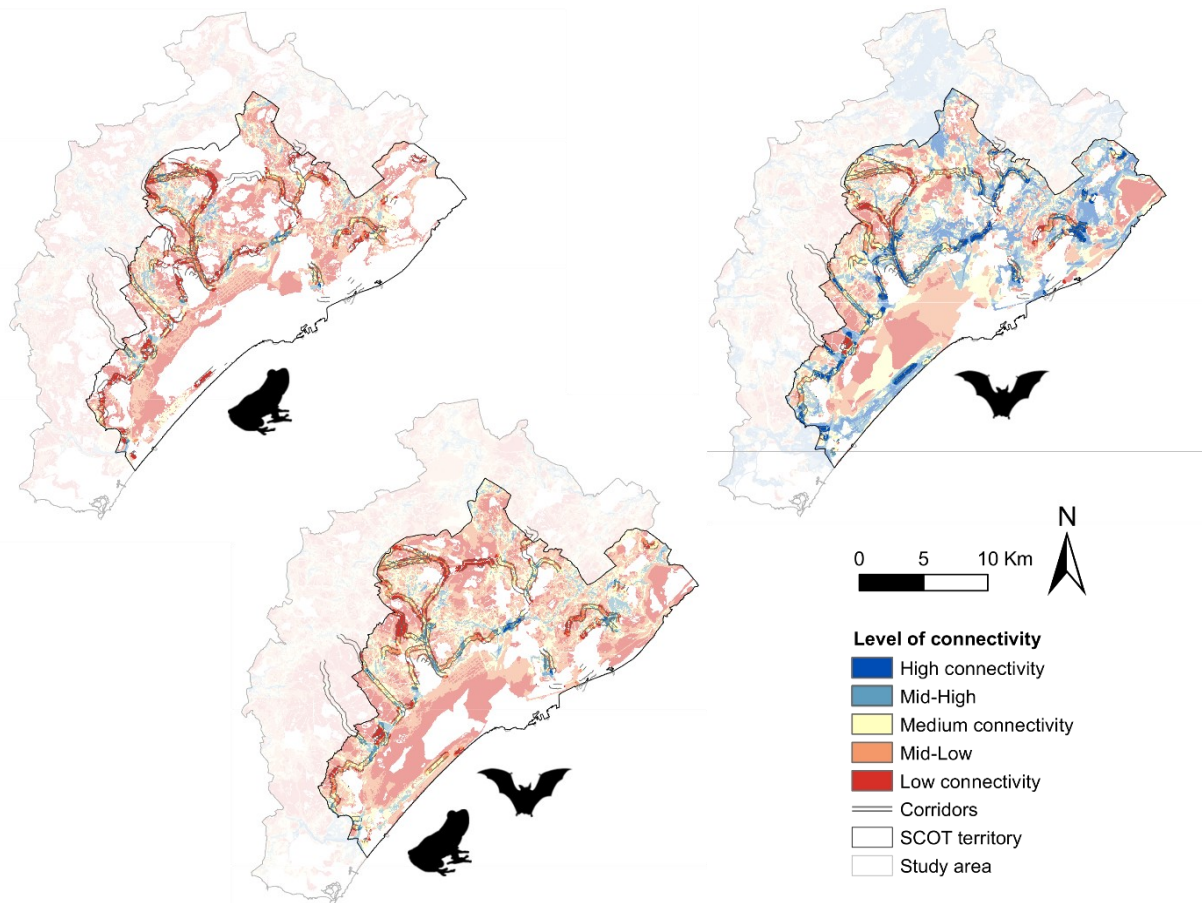

**Fig. 6** Connectivity levels within existing identified corridors, inside a local management scheme (SCOT). White areas correspond to no connectivity level

The urban sprawl projects identified in the SCOT of Thau showed dissimilar impact between dispersal groups (Fig. 7, Online Resource 8). Regarding the low dispersal group, project 1 had the greatest impact with about 9.5 ha mid-low connectivity area. Projects 3 and 2 also showed small impact on areas identified as mid-low for project 3, and low connectivity. Regarding the high dispersal group, project 3 had the greatest impact with about 10 ha of high connectivity area, about 23.5 ha of mid-high connectivity area and about 17.6 ha of medium connectivity area. Projects 1 and 2 also covered mid-high to low connectivity areas for of high dispersal group. Considering all species, projects 1 and 3 were the most unsuitable for connectivity within the study area. On the other hand, the three areas of urban projects at the north-west (4, 5 and 6) of the territory did not seem to be affected by connectivity of the different species according to the analyses (Fig. 7, Online Resource 8).

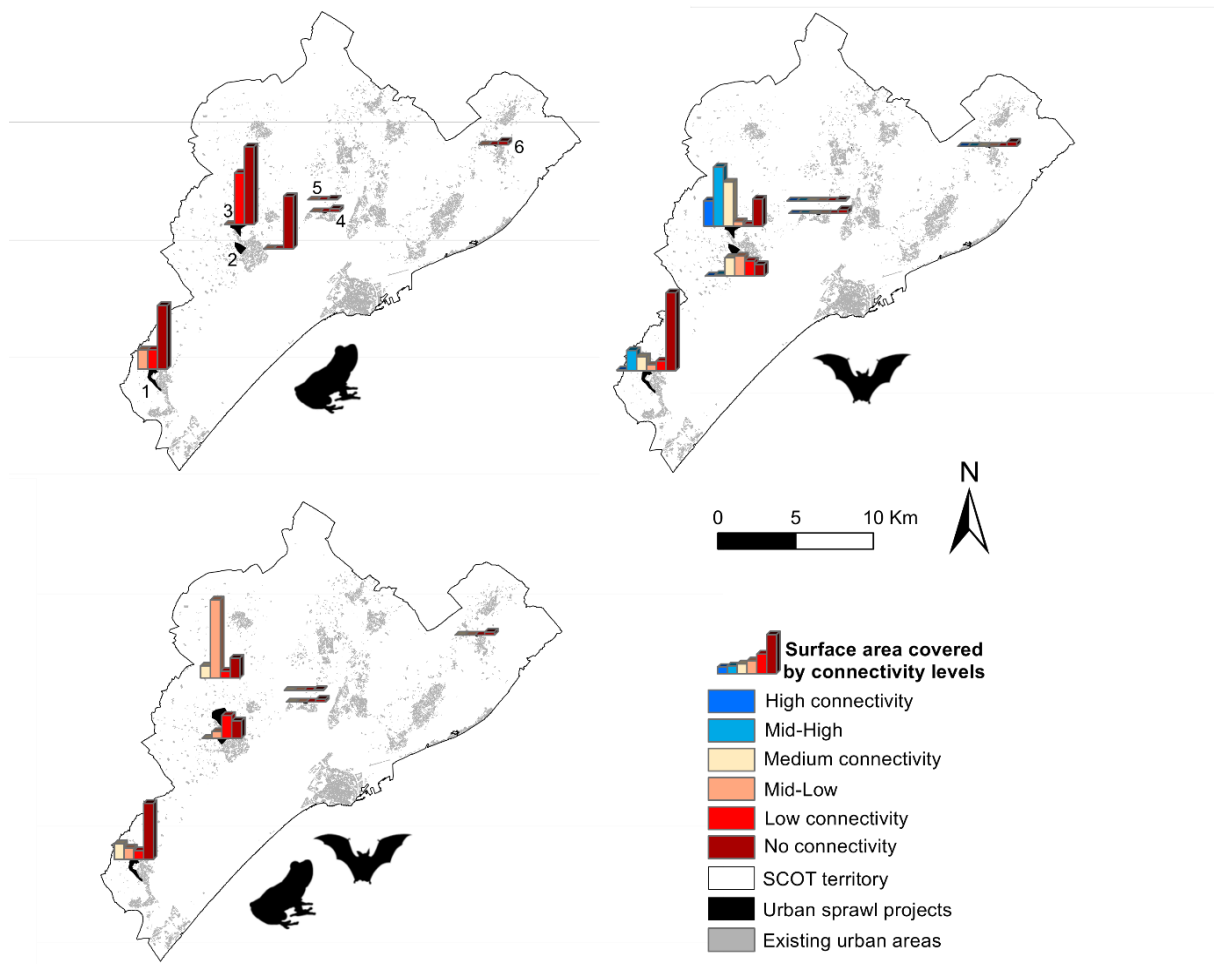

**Fig. 7** Surface area covered by connectivity levels attributed to areas identified for six urban sprawl projects inside a local management scheme (SCOT)

**Discussion**

Accounting for a single group of species with similar dispersal ability can be misleading for area prioritisation. Here, we assessed the effect of dispersal ability on prioritisation of areas based on connectivity levels. Areas identified as corridors for high dispersal species were in most cases identified as no connectivity areas for low dispersal species and conversely. We found important differences in the identification of priority areas between groups with differing dispersal ability. In fact, similarity of predicted connectivity was greater within low or high dispersal single species than between species groups. We found slight similarity of highly connected areas between low dispersal species and high dispersal species. This is the result of little overlap between amphibians' small size habitat patches, especially for *T. marmoratus* and *H. meridionalis*. Conversely, the percentage of area in common was greater for bats due to larger habitat patches, especially for *R. ferrumequinum* and *P. pipistrellus*. The importance of connectivity levels varied between dispersal groups, which may be due to the variety of ecological requirements displayed by the different species.

Long-distance dispersers are commonly thought to be good connectivity surrogates because they allow the identification of larger corridors, covering areas potentially acting as corridors for other species (Diniz et al. 2018). While our results seem in agreement with this statement, the important mismatch between low and high dispersers demonstrates that the omission of low dispersal species can be misleading for conservation planning (Breckheimer et al. 2014). We found substantial differences between average maps of the low and high dispersal groups. For instance, while we identified high connectivity areas covering large surface areas for the high dispersal group, we found not much overlap with the highly connected areas identified for the low dispersal group. We also identified larger areas with no predicted connectivity for the low dispersal group. Furthermore, in our analysis, the corridors already identified in the local urban planning document (SCOT) seemed more suitable to the connectivity of species with high dispersal than those with low dispersal capabilities. Dondina et al. (2020) and Branton and Richardson (2011) have found better performance of low dispersal small-sized species to identify corridors suitable to a broader range of species with different dispersal abilities. This emphasizes the need to account for a variety of dispersal abilities for the identification of ecological corridors.

Confidence maps between groups (here standard deviation) highlighted areas of similarity and divergence regarding classification of connectivity levels according to species and dispersal group. High confidence ensures accounting for most species needs and increases reliability of land planning decisions, while low confidence levels reveal a risk of omitting an important proportion of species. Focusing urban sprawl projects in areas of poor averaged connectivity with high confidence should ensure lesser impact on overall species connectivity. In addition, of the six areas where urban sprawl projects are planned, only the three northernmost areas do not appear to impact areas with important levels of connectivity when considering all species. We recommend the use of uncertainty maps as a tool for enabling to ensuring that the most species are accounted for in urban planning decisions. While ENMs are acknowledged as suitable tools for modelling the distribution of amphibians (*e.g.,* Araújo et al. 2006) and bats (Razgour et al. 2016), one must be aware that such assessments are limited by several biases inherent to ENMs, as reviewed by Jarnevich et al. (2015). For instance, species occurrences are likely to be biased towards the most accessible or protected areas (Dubos et al. 2021), which we have accounted in our methodology by selecting PsA with the same bias as occurrence data. Our results have, however, unavoidably been influenced by modelling choices; for instance, by the approach of connectivity analysis used and the distance to simulate connectivity (Zeller et al. 2018; Diniz et al. 2020). Besides, while outputs of ENMs are commonly used to determine resistance surfaces, it has been shown to produce different results than when resistance is based on land use, like in amphibians (Godet and Clauzel 2021).

We identified a gap in the level of protection of low dispersal species, particularly in areas with an important connectivity level. Indeed, we found high connectivity levels for both low and high dispersal groups within the nature reserve while elevated connectivity was identified in the high dispersal group only in the area of biotope protection. Whilst we found more important connectivity levels in protected areas than over the entire study area for each dispersal group, this proved not to be the case for Natura 2000 areas (SPAs and SCIs) which did not encompass greater levels of connectivity than the entire study area. These results are consistent with the protected areas partially designed to protect the study species of the two dispersal groups. While SPAs are designed for the conservation of bird species identified in the Birds Directive and their ability to act as umbrellas seemed limited in the light of our results. Furthermore, several studies identified differences in the representation of taxa within Natura 2000 sites (*e.g.,* Trochet and Schmeller 2013; Kukkala et al. 2016; Zisenis 2017). While SCIs in the south and the east of the study area were identified mainly at the sea pond level, different species of bats and amphibians were listed within the SCI site located in the northeast of the study area. The connectivity levels of high dispersal species seemed better represented within this site than those of low dispersal species. Our results agree with the observations of Abellán and Sánchez-Fernández (2015) who found that narrow-ranging species were less covered by Natura 2000 sites. In addition, while some authors advocate for the update of the species listed in the Habitats and Birds Directives (Maiorano et al. 2015; Hermoso et al. 2019), one should recall that Natura 2000 sites are not strictly protected although they are targeted by conservation objectives, implying a greater impact of human activities on species than in protected areas having different status.

All in all, we recommend grouping species by taxa, similar dispersal mode and abilities along with different habitat and resource characteristics, in order to cover a large range of connectivity requirements. For instance, Lechner et al. (2017) have developed an approach called "dispersal guild", where they grouped mammals by similar dispersal behaviour and habitat characteristics, which allowed accounting for the multispecies dispersal needs in conservation planning. Herrera et al. (2018) have also assessed multispecies connectivity for birds based on guild-specific responses to environmental variables allowing the optimisation of restoration actions. We suggest replicating analogous methodology for various taxa. An accurate representation of multi-species connectivity would require considering a large number of species, for instance Kujala et al. (2018) recommended to account for at least 50 species to increase the stability of results. However, as it may be challenging to access such a large amount of data particularly in small territories; our methodology stands as a compromise promising an operational and optimised solution to guide conservation decisions within the context of territorial planning aiming conservation. In the face of current land use pressures and climate change it is critical to identify ecological networks while safeguarding connectivity for species with heterogeneous dispersal abilities (Keeley et al. 2018; Costanza and Terando 2019). Indeed, biodiversity is subject to habitat loss and fragmentation exacerbated by climate change at a global scale (Segan et al. 2016), which are forecast to increase in severity over future decades (Newbold 2018). This forecast includes common species that may become threatened in the near future (Inger et al. 2015; Préau et al. 2021). As a result, connectivity assessment should not focus only on rare, charismatic, or protected species, but should also consider common species. In order to ensure anticipated management and conservation of overall biodiversity in line with recent conservation guidelines, which now advocate the integration of phylogenetic and community metrics (*e.g.*, species richness, rarity, originality, functional diversity; Tucker et al. 2017; Kondratyeva et al. 2019), diversity in dispersal capacity must be accounted for in the context of programmes aiming to maintain, protect or restore habitat connectivity.

**Acknowledgements**

We thank The French League for the Protection of Birds (LPO Occitanie) and the Chiroptera group of Languedoc Roussillon (GCLR) for sharing their data and expertise on the species. We thank the "Syndicat mixte du bassin de Thau" and, particularly Camille Pfleger and David Cottalorda for their useful insight into planning needs for the region. We also thank Pierre Maurel for sharing his expertise on territorial requirements in the area. We thank Marcus Cork-Keeling for the English language revision. This study was partly supported by the IMAGINE project (ERANET BIODIVERSA EU Program). The work of ML was supported by a grant from the French National Research Agency (project NetCost, ANR-17-CE03-0003 grant).

**Conflicts of interest/Competing interests** The authors declare no conflict of interest.

**Code availability** Codes for classification method and computation of similarity metric are available at: https://gitlab.com/maximelenormand/mismatch-connectivity

# References

Abellán P, Sánchez-Fernández D (2015) A gap analysis comparing the effectiveness of Natura 2000 and national protected area networks in representing European amphibians and reptiles. Biodivers Conserv 24:1377–1390. https://doi.org/10.1007/s10531-015-0862-3

Adriaensen F, Chardon JP, De Blust G, Swinnen E, Villalba S, Gulinck H, Matthysen E (2003) The application of 'least-cost' modelling as a functional landscape model. Landsc Urban Plan 64:233–247. https://doi.org/10.1016/S0169-2046(02)00242-6

Alagador D, Trivino M, Cerdeira JO, Brás R, Cabeza M, Araújo, MB (2012) Linking like with like: optimising connectivity between environmentally-similar habitats. Landsc Ecol 27: 291–301. https://doi.org/10.1007/s10980-012-9704-9

Albert CH, Rayfield B, Dumitru M, Gonzalez A (2017) Applying network theory to prioritize multispecies habitat networks that are robust to climate and land-use change. Conserv Biol 31:1383–1396. https://doi.org/10.1111/cobi.12943

Allan E, Manning P, Alt F, et al (2015) Land use intensification alters ecosystem multifunctionality via loss of biodiversity and changes to functional composition. Ecol Lett 18:834–843. https://doi.org/10.1111/ele.12469

Allouche O, Tsoar A, Kadmon R (2006) Assessing the accuracy of species distribution models: prevalence, kappa and the true skill statistic (TSS). J Appl Ecol 43:1223–1232. https://doi.org/10.1111/j.1365-2664.2006.01214.x

Araújo MB, Thuiller W, Pearson RG (2006) Climate warming and the decline of amphibians and reptiles in Europe. J Biogeogr 33:1712–1728. https://doi.org/10.1111/j.1365-2699.2006.01482.x

Baguette M, Blanchet S, Legrand D, Stevens VM, Turlure C (2013) Individual dispersal, landscape connectivity and ecological networks. Biol Rev 88:310–326. https://doi.org/10.1111/brv.12000

Bassolas A, Barbosa-Filho H, Dickinson B, et al (2019) Hierarchical organization of urban mobility and its connection with city livability. Nat Commun 10:4817. https://doi.org/10.1038/s41467-019-12809-y

Biondi E, Casavecchia S, Pesaresi S, Zivkovic L (2012) Natura 2000 and the Pan-European Ecological Network: a new methodology for data integration. Biodivers Conserv 21:1741–1754. https://doi.org/10.1007/s10531-012-0273-7

Blasi C, Zavattero L, Marignani M, Smiraglia D, Copiz R, Rosati L, Del Vico E (2008) The concept of land ecological network and its design using a land unit approach. Plant Biosyst - Int J Deal Asp Plant Biol 142:540–549. https://doi.org/10.1080/11263500802410892

Boyce MS, Vernier PR, Nielsen SE, Schmiegelow FKA (2002) Evaluating resource selection functions. Ecol Model 157:281–300. https://doi.org/10.1016/S0304-3800(02)00200-4

Branton M, Richardson JS (2011) Assessing the Value of the Umbrella-Species Concept for Conservation Planning with Meta-Analysis. Conserv Biol 25:9–20. https://doi.org/10.1111/j.1523-1739.2010.01606.x

Brás R, Cerdeira JO, Alagador D, Araújo MB (2013) Linking habitats for multiple species. Environmental Modelling & Software, 40 : 336–339. https://doi.org/10.1016/j.envsoft.2012.08.001

Breckheimer I, Haddad NM, Morris WF, et al (2014) Defining and Evaluating the Umbrella Species Concept for Conserving and Restoring Landscape Connectivity. Conserv Biol 28:1584–1593. https://doi.org/10.1111/cobi.12362

Brun P, Thuiller W, Chauvier Y, et al (2020) Model complexity affects species distribution projections under climate change. J Biogeogr 47: 130–142. https://doi.org/10.1111/jbi.13734

Buhk C, Alt M, Steinbauer MJ, Beierkuhnlein C, Warren SD, Jentsch A (2017) Homogenizing and diversifying effects of intensive agricultural land-use on plant species beta diversity in Central Europe — A call to adapt our conservation measures. Sci Total Environ 576:225–233. https://doi.org/10.1016/j.scitotenv.2016.10.106

Capinha C, Larson ER, Tricarico E, Olden JD, Gherardi F (2013) Effects of climate change, invasive species, and disease on the distribution of native European crayfishes. Conserv Biol 27:731–40. https://doi.org/10.1111/cobi.12043

Chatterjee S, Hadi AS (2006) Regression analysis by example. John Wiley and Sons.

Correa Ayram CA, Mendoza ME, Etter A, Salicrup DRP (2016) Habitat connectivity in biodiversity conservation: A review of recent studies and applications. Prog Phys Geogr 40:7–37. https://doi.org/10.1177/0309133315598713

Costanza JK, Terando AJ (2019) Landscape Connectivity Planning for Adaptation to Future Climate and Land-Use Change. Curr Landsc Ecol Rep 4:1–13. https://doi.org/10.1007/s40823-019-0035-2

Decout S, Manel S, Miaud C, Luque S (2012) Integrative approach for landscape-based graph connectivity analysis: a case study with the common frog (*Rana temporaria*) in human-dominated landscapes. Landsc Ecol 27:267–279. https://doi.org/10.1007/s10980-011-9694-z

Dickson BG, Albano CM, Anantharaman R, et al (2019) Circuit-theory applications to connectivity science and conservation. Conserv Biol 33:239–249. https://doi.org/10.1111/cobi.13230

Diniz MF, Cushman SA, Machado RB, De Marco Júnior P (2020) Landscape connectivity modeling from the perspective of animal dispersal. Landsc Ecol 35:41–58. https://doi.org/10.1007/s10980-019-00935-3

Diniz MF, Machado RB, Bispo AA, De M. Júnior P (2018) Can we face different types of storms under the same umbrella? Efficiency and consistency of connectivity umbrellas across different patchy landscape patterns. Landsc Ecol 33:1911–1923. https://doi.org/10.1007/s10980-018-0720-2

Dondina O, Orioli V, Chiatante G, Bani L (2020) Practical insights to select focal species and design priority areas for conservation. Ecol Indic 108:105767. https://doi.org/10.1016/j.ecolind.2019.105767

Dubos N, Préau C, Lenormand M, et al (2021) Assessing the effect of sampling bias correction in species distribution models. arXiv preprint arXiv:2103.07107.

Duflot R, Avon C, Roche P, Bergès L (2018) Combining habitat suitability models and spatial graphs for more effective landscape conservation planning: An applied methodological framework and a species case study. J Nat Conserv 46:38–47. https://doi.org/10.1016/j.jnc.2018.08.005

European Commission Natura 2000. https://ec.europa.eu/environment/nature/natura2000/index_en.htm. Accessed 14 April 2021

Fitzpatrick MC, Gotelli NJ, Ellison AM (2013) MaxEnt versus MaxLike: empirical comparisons with ant species distributions. Ecosphere 4:art55. https://doi.org/10.1890/ES13-00066.1

Fonte LM, Mayer M, Lötters S (2019) Long-distance dispersal in amphibians. Front Biogeogr 11:e44577. https://doi.org/10.21425/F5FBG44577

Godet C, Clauzel C (2021) Comparison of landscape graph modelling methods for analysing pond network connectivity. Landsc Ecol 36:735–748. https://doi.org/10.1007/s10980-020-01164-9

Hanley JA, McNeil BJ (1982) The meaning and use of the area under a receiver operating characteristic (ROC) curve. Radiology 143:29–36. https://doi.org/10.1148/radiology.143.1.7063747

Hermoso V, Morán-Ordóñez A, Canessa S, Brotons L (2019) Realising the potential of Natura 2000 to achieve EU conservation goals as 2020 approaches. Sci Rep 9:16087. https://doi.org/10.1038/s41598-019-52625-4

Herrera JM, Alagador D, Salgueiro P, Mira A (2018) A distribution-oriented approach to support landscape connectivity for ecologically distinct bird species. PLoS ONE 13: e0194848. https://doi.org/10.1371/journal.pone.0194848

Hirzel AH, Le Lay G, Helfer V, Randin C, Guisan A (2006) Evaluating the ability of habitat suitability models to predict species presences. Predict Species Distrib 199:142–152. https://doi.org/10.1016/j.ecolmodel.2006.05.017

Inger R, Gregory R, Duffy JP, Stott J, Voříšek P, Gaston KJ (2015) Common European birds are declining rapidly while less abundant species' numbers are rising. Ecol Lett 18:28–36. https://doi.org/10.1111/ele.12387

Jalkanen J, Toivonen T, Moilanen A (2020) Identification of ecological networks for land-use planning with spatial conservation prioritization. Landsc Ecol 35:353–371. https://doi.org/10.1007/s10980-019-00950-4

Jarnevich CS, Stohlgren TJ, Kumar S, Morisette JT, Holcombe TR (2015) Caveats for correlative species distribution modeling. Ecol Inform 29:6–15. https://doi.org/10.1016/j.ecoinf.2015.06.007

Kaim D, Ziółkowska E, Szwagrzyk M, Price B, Kozak J (2019) Impact of Future Land Use Change on Large Carnivores Connectivity in the Polish Carpathians. Land 8:8. https://doi.org/10.3390/land8010008

Keeley ATH, Ackerly DD, Cameron DR, et al (2018) New concepts, models, and assessments of climate-wise connectivity. Environ Res Lett 13:073002. https://doi.org/10.1088/1748-9326/aacb85

Keeley ATH, Beier P, Creech T, Jones K, Jongman RHG, Stonecipher G, Tabor GM (2019) Thirty years of connectivity conservation planning: an assessment of factors influencing plan implementation. Environ Res Lett 14:103001. https://doi.org/10.1088/1748-9326/ab3234

Kondratyeva A, Grandcolas P, Pavoine S (2019) Reconciling the concepts and measures of diversity, rarity and originality in ecology and evolution. Biol Rev 94:1317–1337. https://doi.org/10.1111/brv.12504

Kujala H, Moilanen A, Gordon A (2018) Spatial characteristics of species distributions as drivers in conservation prioritization. Methods Ecol Evol 9:1121–1132. https://doi.org/10.1111/2041-210X.12939

Kukkala AS, Arponen A, Maiorano L, et al (2016) Matches and mismatches between national and EU-wide priorities: Examining the Natura 2000 network in vertebrate species conservation. Biol Conserv 198:193–201. https://doi.org/10.1016/j.biocon.2016.04.016

Lambeck RJ (1997) Focal Species: A Multi-Species Umbrella for Nature Conservation. Conserv Biol 11:849–856. https://doi.org/10.1046/j.1523-1739.1997.96319.x

Le Roux M, Redon M, Archaux F, Long J, Vincent S, Luque S (2017) Conservation planning with spatially explicit models: a case for horseshoe bats in complex mountain landscapes. Landsc Ecol 32:1005–1021. https://doi.org/10.1007/s10980-017-0505-z

Lechner AM, Sprod D, Carter O, Lefroy EC (2017) Characterising landscape connectivity for conservation planning using a dispersal guild approach. Landsc Ecol 32:99–113. https://doi.org/10.1007/s10980-016-0431-5

Lindenmayer DB, Manning AD, Smith PL, Possingham HP, Fischer J, Oliver I, McCarthy MA (2002) The Focal-Species Approach and Landscape Restoration: a Critique. Conserv Biol 16:338–345. https://doi.org/10.1046/j.1523-1739.2002.00450.x

Lizée M-H, Manel S, Mauffrey J-F, Tatoni T, Deschamps-Cottin M (2012) Matrix configuration and patch isolation influences override the species–area relationship for urban butterfly communities. Landsc Ecol 27:159–169. https://doi.org/10.1007/s10980-011-9651-x

Louail T, Lenormand M, Cantu Ros OG, et al (2014) From mobile phone data to the spatial structure of cities. Sci Rep 4:5276. https://doi.org/10.1038/srep05276

Luque S, Saura S, Fortin M-J (2012) Landscape connectivity analysis for conservation: insights from combining new methods with ecological and genetic data. Landsc Ecol 27:153–157. https://doi.org/10.1007/s10980-011-9700-5

Maiorano L, Amori G, Montemaggiori A, Rondinini C, Santini L, Saura S, Boitani L (2015) On how much biodiversity is covered in Europe by national protected areas and by the Natura 2000 network: insights from terrestrial vertebrates. Conserv Biol 29:986–995. https://doi.org/10.1111/cobi.12535

McRae B, Kavanagh D (2011) Linkage Mapper Connectivity Analysis Software. The Nature Conservancy, Seattle, WA

McRae BH, Dickson BG, Keitt TH, Shah VB (2008) Using Circuit Theory to model connectivity in ecology, evolution, and conservation. Ecology 89:2712–2724. https://doi.org/10.1890/07-1861.1

Meurant M, Gonzalez A, Doxa A, Albert CH (2018) Selecting surrogate species for connectivity conservation. Biol Conserv 227:326–334. https://doi.org/10.1016/j.biocon.2018.09.028

Ministère de la Cohésion des territoires et des Relations avec les collectivités territoriales Le SCOT: Un projet stratégique partagé pour l'aménagement d'un territoire. https://www.cohesion-territoires.gouv.fr/le-scot-un-projet-strategique-partage-pour-lamenagement-dun-territoire. Accessed 14 April 2021

Naimi B, Hamm NAS, Groen TA, Skidmore AK, Toxopeus AG (2014) Where is positional uncertainty a problem for species distribution modelling? Ecography 37: 191–203. https://doi.org/ 10.1111/j.1600-0587.2013.00205.x

Naimi B, Araújo MB (2016) sdm: a reproducible and extensible R platform for species distribution modelling. Ecography 39: 368–375. https://doi.org/10.1111/ecog.01881

Newbold T (2018) Future effects of climate and land-use change on terrestrial vertebrate community diversity under different scenarios. Proc R Soc B Biol Sci 285:20180792. https://doi.org/10.1098/rspb.2018.0792

Pearson RG, Raxworthy CJ, Nakamura M, Townsend Peterson A (2007) Predicting species distributions from small numbers of occurrence records: a test case using cryptic geckos in Madagascar. J Biogeogr 34:102–117. https://doi.org/10.1111/j.1365-2699.2006.01594.x

Phillips SJ, Dudík M, Elith J, Graham CH, Lehmann A, Leathwick J, Ferrier S (2009) Sample selection bias and presence-only distribution models: implications for background and pseudo-absence data. Ecol Appl 19:181–197. https://doi.org/10.1890/07-2153.1

Préau C, Bertrand R, Sellier Y, Grandjean F, Isselin-Nondedeu F (2021) Climate change would prevail over land use change in shaping the future distribution of *Triturus marmoratus* in France. Anim. Conserv. https://doi.org/10.1111/acv.12733

R Core Team (2020) R: A language and environment for statistical computing. R Foundation for Statistical Computing, Vienna, Austria

Ramirez-Reyes C, Bateman BL, Radeloff VC (2016) Effects of habitat suitability and minimum patch size thresholds on the assessment of landscape connectivity for jaguars in the Sierra Gorda, Mexico. Biol Conserv 204:296–305. https://doi.org/10.1016/j.biocon.2016.10.020

Razgour O, Rebelo H, Di Febbraro M, Russo D (2016) Painting maps with bats: species distribution modelling in bat research and conservation. Hystrix Ital J Mammal 27:1. https://doi.org/10.4404/hystrix-27.1-11753

Richardson JL (2012) Divergent landscape effects on population connectivity in two co-occurring amphibian species. Mol Ecol 21:4437–4451. https://doi.org/10.1111/j.1365-294X.2012.05708.x

Rudnick D, Ryan SJ, Beier P, et al (2012) The role of landscape connectivity in planning and implementing conservation and restoration priorities. Issues in Ecology 16:1-20.

Runge CA, Martin TG, Possingham HP, Willis SG, Fuller RA (2014) Conserving mobile species. Front Ecol Environ 12:395–402. https://doi.org/10.1890/130237

Santini L, Di Marco M, Visconti P, Baisero D, Boitani L, Rondinini C (2013) Ecological correlates of dispersal distance in terrestrial mammals. Hystrix Ital J Mammal 24:181–186. https://doi.org/10.4404/hystrix-24.2-8746

Segan DB, Murray KA, Watson JEM (2016) A global assessment of current and future biodiversity vulnerability to habitat loss–climate change interactions. Glob Ecol Conserv 5:12–21. https://doi.org/10.1016/j.gecco.2015.11.002

Stevens VM, Verkenne C, Vandewoestijne S, Wesselingh RA, Baguette M (2006) Gene flow and functional connectivity in the natterjack toad. Mol Ecol 15:2333–2344. https://doi.org/10.1111/j.1365-294X.2006.02936.x

Taylor PD, Fahrig L, Henein K, Merriam G (1993) Connectivity Is a Vital Element of Landscape Structure. Oikos 68:571–573. https://doi.org/10.2307/3544927

Thuiller W, Lafourcade B, Engler R, Araújo MB (2009) BIOMOD – a platform for ensemble forecasting of species distributions. Ecography 32:369–373. https://doi.org/10.1111/j.1600-0587.2008.05742.x

Tischendorf L, Fahrig L (2000) How should we measure landscape connectivity? Landsc Ecol 15:633–641. https://doi.org/10.1023/A:1008177324187

Tournayre O, Pons J-B, Leuchtmann M, et al (2019) Integrating population genetics to define conservation units from the core to the edge of Rhinolophus ferrumequinum western range. Ecol Evol 9:12272–12290. https://doi.org/10.1002/ece3.5714

Trochet A, Schmeller DS (2013) Effectiveness of the Natura 2000 network to cover threatened species. Nat Conserv 4:35–53. https://doi.org/10.3897/natureconservation.4.3626

Tucker CM, Cadotte MW, Carvalho SB, et al (2017) A guide to phylogenetic metrics for conservation, community ecology and macroecology. Biol Rev 92:698–715. https://doi.org/10.1111/brv.12252

Urban D, Keitt T (2001) Landscape Connectivity: A Graph-Theoretic Perspective. Ecology 82:1205–1218. https://doi.org/10.2307/2679983

WHCWG (Washington Wildlife Habitat Connectivity Working Group) (2010) Washington Connected Landscapes Project: Statewide Analysis. https://waconnected.org/statewide-analysis/. Accessed 12 March 2021

WWF (2020) Living Planet Report 2020 - Bending the curve of biodiversity loss. WWF, Gland, Switzerland.

Zeller KA, Jennings MK, Vickers TW, Ernest HB, Cushman SA, Boyce WM (2018) Are all data types and connectivity models created equal? Validating common connectivity approaches with dispersal data. Divers Distrib 24:868–879. https://doi.org/10.1111/ddi.12742

Zeller KA, McGarigal K, Whiteley AR (2012) Estimating landscape resistance to movement: a review. Landsc Ecol 27:777–797. https://doi.org/10.1007/s10980-012-9737-0

Zisenis M (2017) Is the Natura 2000 network of the European Union the key land use policy tool for preserving Europe's biodiversity heritage? Land Use Policy 69:408–416. https://doi.org/10.1016/j.landusepol.2017.09.045

Online Resource 1: Species information and results of ecological niche modelling

**Table 1** Species information and results of ENM evaluation

| Group | Species | Distance used in analysis | References | Number of occurrences | AUC-ROC | TSS | Boyce Index |
|---|---|---|---|---|---|---|---|
| Low dispersal group | *Triturus marmoratus* | 10 km | Fonte et al. 2019 | 64 | 0.994 | 0.953 | 0.76 |
| | *Hyla meridionalis* | 10 km | Fonte et al. 2019 | 233 | 0.978 | 0.845 | 0.966 |
| | *Epidalea calamita* | 10 km | Fonte et al. 2019 | 89 | 0.994 | 0.91 | 0.968 |
| | *Pelodytes punctatus* | 10 km | Fonte et al. 2019 | 153 | 0.977 | 0.85 | 0.978 |
| High dispersal group | *Rhinolophus ferrumequinum* | 500 km | Tournayre et al. 2019 | 44 | 0.983 | 0.882 | 0.872 |
| | *Miniopterus schreibersii* | 500 km | Garin et al. 2008; Ramos Pereira et al. 2009 | 81 | 0.951 | 0.992 | 0.969 |
| | *Nyctalus leisleri* | 2000 km | Furmankiewicz and Kucharska 2009 | 77 | 0.996 | 0.948 | 0.967 |
| | *Pipistrellus pipistrellus* | 2000 km | Furmankiewicz and Kucharska 2009 | 140 | 0.943 | 0.753 | 0.899 |

Fonte LM, Mayer M, Lötters S (2019) Long-distance dispersal in amphibians. Front Biogeogr 11:e44577

Furmankiewicz J, Kucharska M (2009) Migration of Bats along a Large River Valley in Southwestern Poland. J Mammal 90:1310–1317. https://doi.org/10.1644/09-MAMM-S-099R1.1

Garin I, Aihartza J, Agirre-Mendi PT, et al (2008) Seasonal movements of the Schreibers' bat, Miniopterus schreibersii, in the northern Iberian Peninsula. Ital J Zool 75:263–270. https://doi.org/10.1080/11250000801927850

Ramos Pereira MJ, Salgueiro P, Rodrigues L, et al (2009) Population Structure of a Cave-Dwelling Bat, Miniopterus schreibersii: Does It Reflect History and Social Organization? J Hered 100:533–544. https://doi.org/10.1093/jhered/esp032

Tournayre O, Pons J-B, Leuchtmann M, et al (2019) Integrating population genetics to define conservation units from the core to the edge of Rhinolophus ferrumequinum western range. Ecol Evol 9:12272–12290. https://doi.org/10.1002/ece3.5714

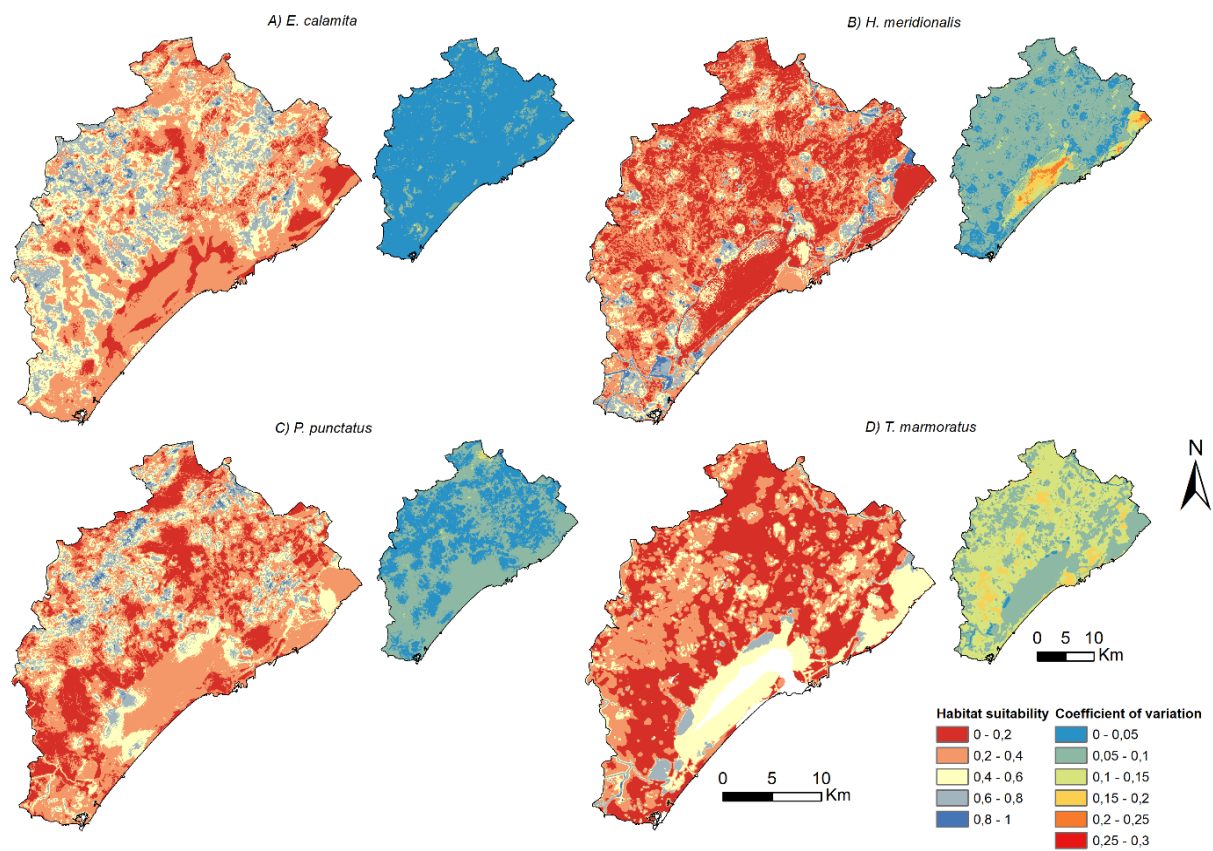

**Fig. 1** Results of ENM for amphibian species

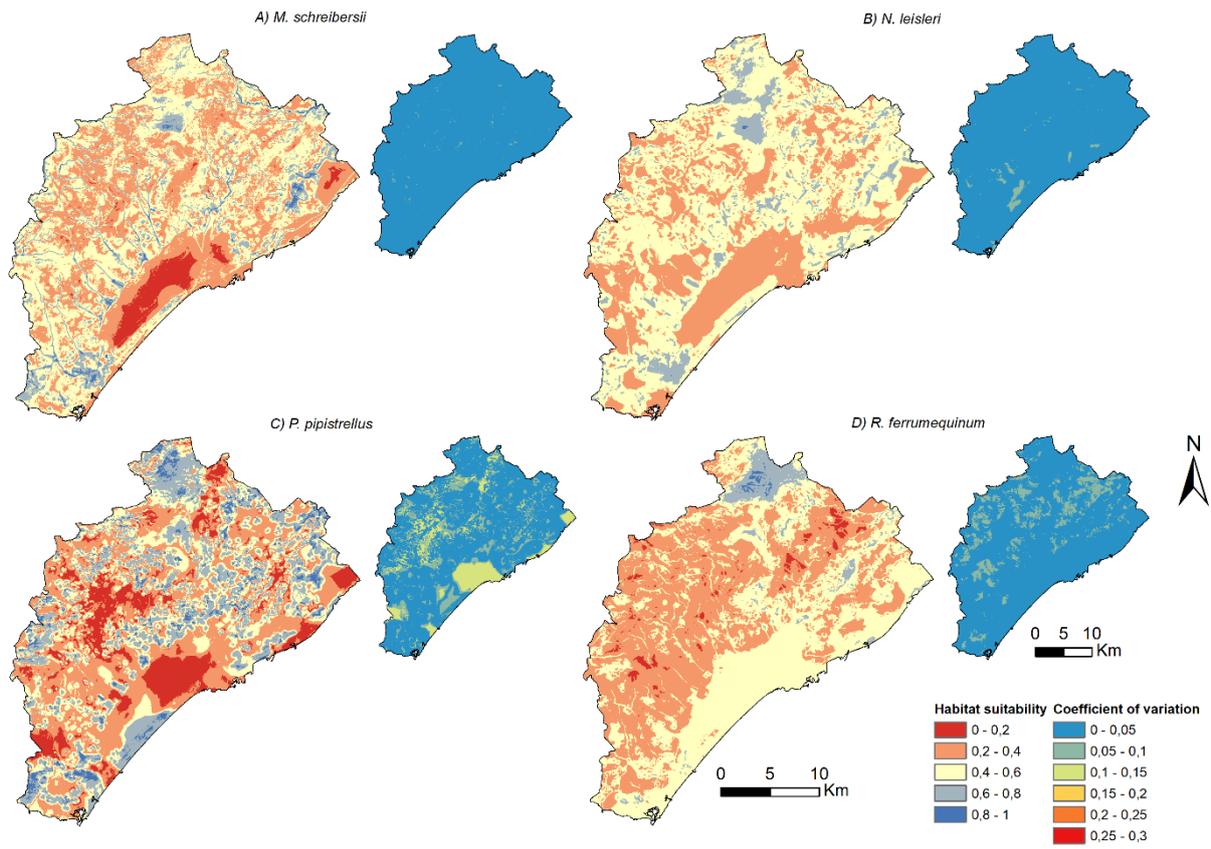

**Fig. 2** Results of ENM for bat species

Online Resource 2: Environmental data

**Table 2** Source of environmental data

| Name | Source |
|---|---|
| Elevation | IGN - MNT - RGE ALTI |
| Built density | BD TOPO IGN |
| Distance to roads | BD TOPO IGN |
| Distance to urban areas (with 25m buffer) | CLC 231, 242, 244 |
| Distance to water bodies | BD TOPO IGN |
| Distance to water courses | BD TOPO IGN |
| Maximum annual difference vegetation index (NDVI) | THEIA Land |
| Distance to deciduous forest edges | BD forets IGN |
| Distance to coniferous forest edges | BD forets IGN |
| Distance to mixed forest edges | BD forets IGN |
| Distance to scrubland | BD forets IGN |
| Distance to intensive agricultural areas | CLC 211, 221, 223 |
| Distance to non-intensive agricultural areas | CLC 111, 112, 121, 123, 131, 141, 142 |

Online Resource 3: Method for classification of levels of connectivity from Pinchpoint mapper outputs

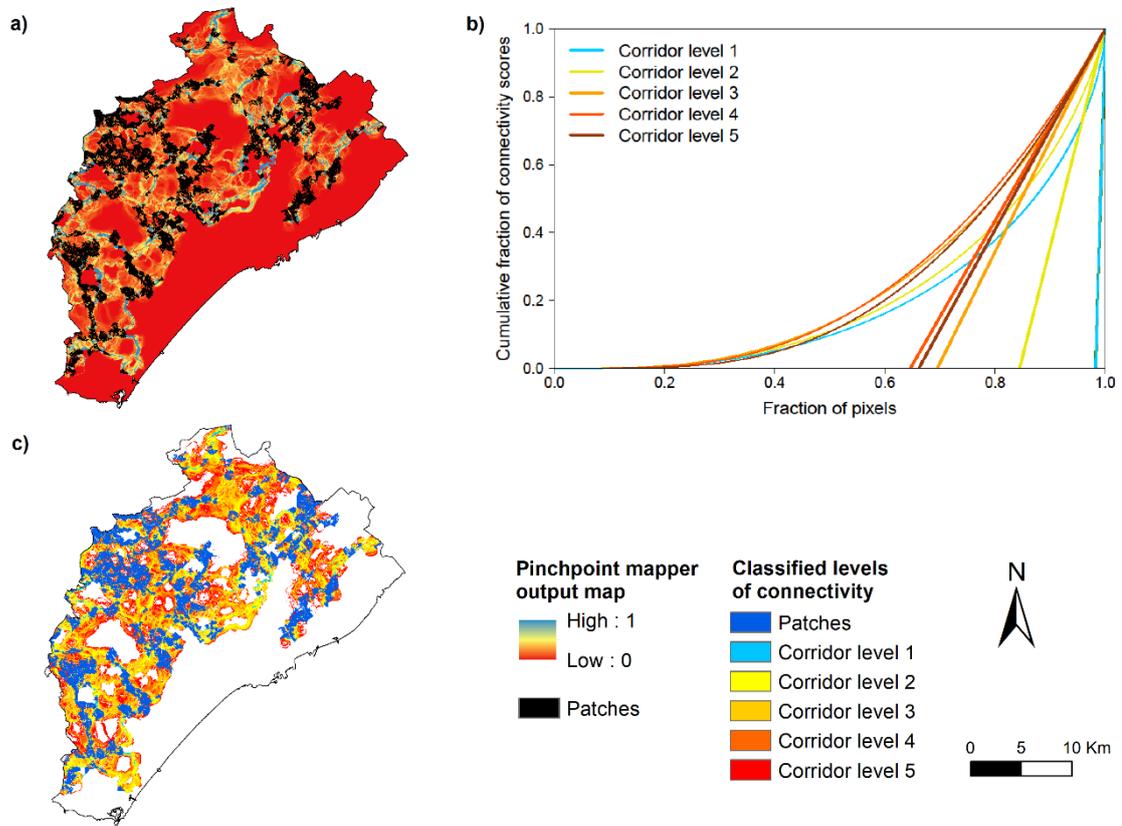

**Fig. 1** a) Example of output map of Pinchpoint mapper depicting connectivity values (0 to 1), and patches. b) Iterative hotspot level calculation using the Loubar method (Louail et al. 2014). The Loubar method is first applied on the Lorentz curve of the full distribution of connectivity scores (green curve) to identify the first level of corridors (based on the intersection between the tangent of the curve at point (1,1) and the x-axis). Then the Loubar method is iteratively applied on the distribution of connectivity scores without considering pixels identified at previous steps. c) Output maps of classified levels of connectivity

Online Resource 4: Computation of similarity metric

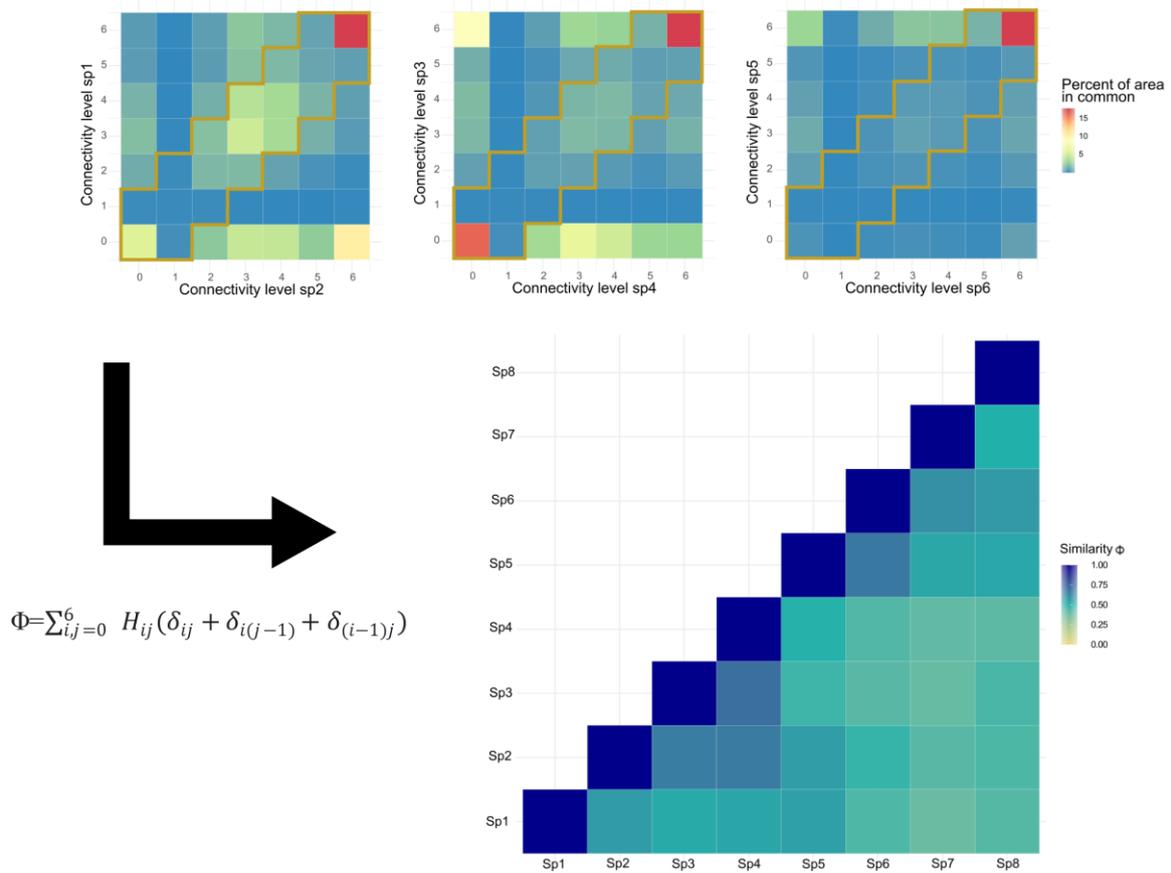

$$\Phi = \sum_{i,j=0}^{6} H_{ij} (\delta_{ij} + \delta_{i(j-1)} + \delta_{(i-1)j})$$

**Fig. 1** Graphical representation of the similarity metric Φ as the tri-diagonal trace of H where $\delta_{ij}$ is the Kronecker delta (Bassolas et al. 2019) and H is a matrix with element $H_{ij}$ corresponding to the number of shared pixels values between the resulting levels of connectivity assigned for each species (here represented in percent of area in common)

Online Resource 5: Pairwise identification of percentage of common area within connectivity levels between low dispersal species, high dispersal species and all species

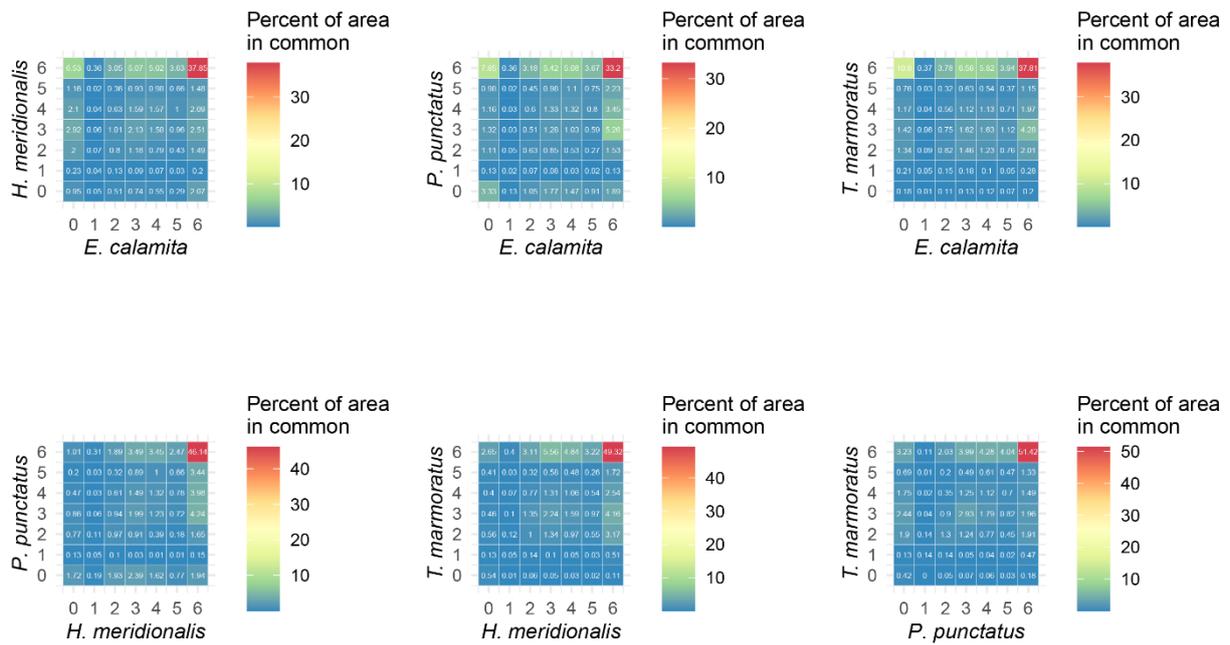

**Fig. 1** Pairwise identification of percentage of common area with connectivity levels between low dispersal species

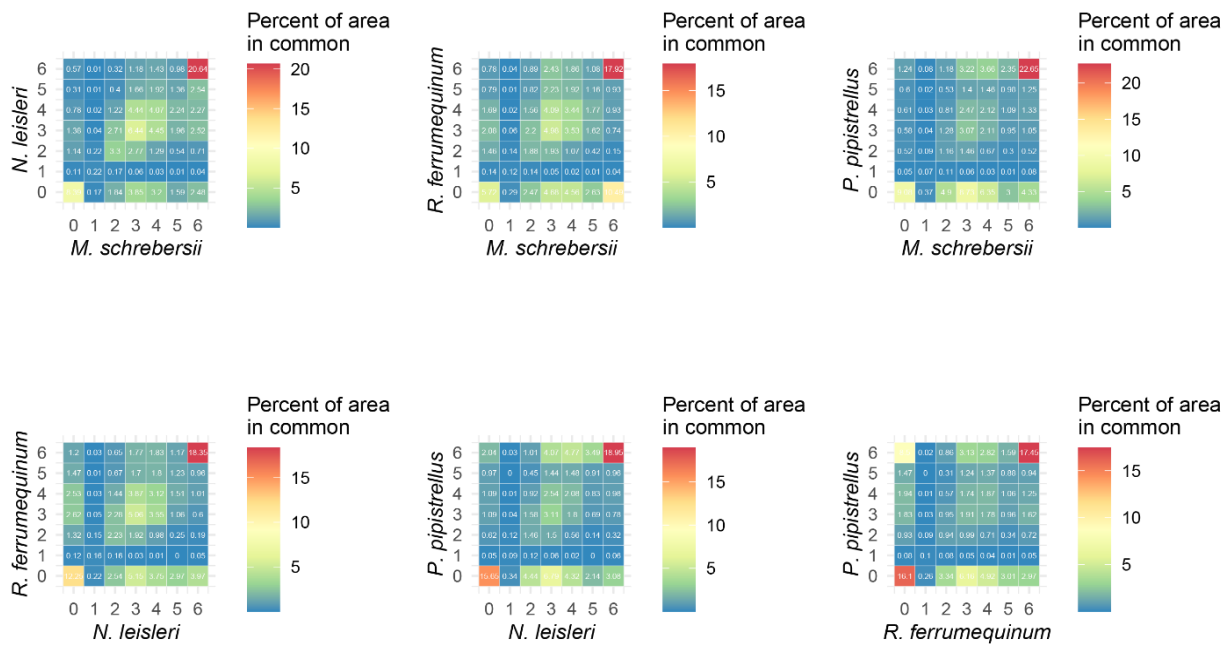

**Fig. 2** Pairwise identification of percentage of common area with connectivity levels between high dispersal species

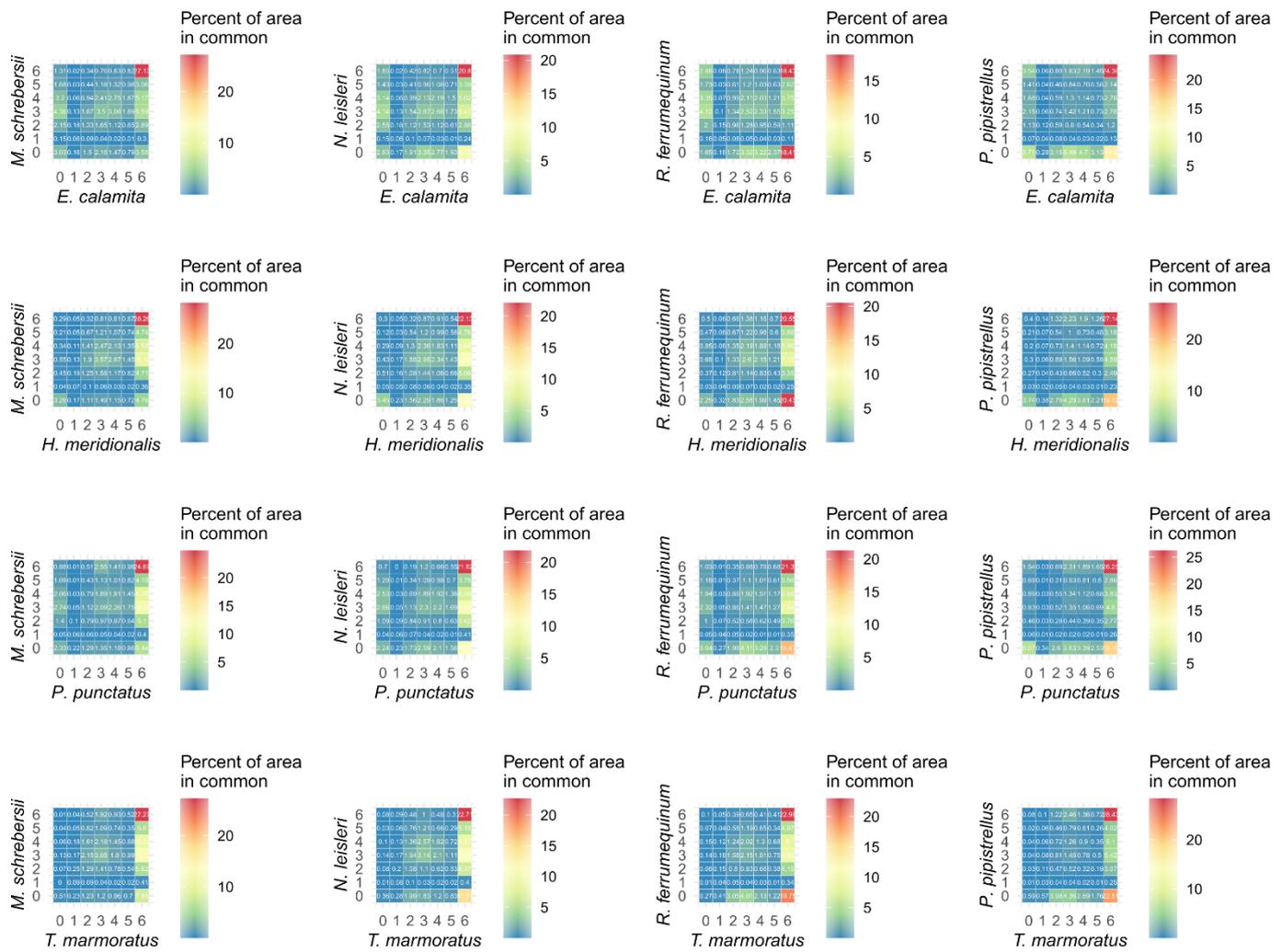

**Fig. 3** Pairwise identification of percentage of common area with connectivity levels between all species

Online Resource 6: Single maps of connectivity levels and average map for low and high dispersal groups

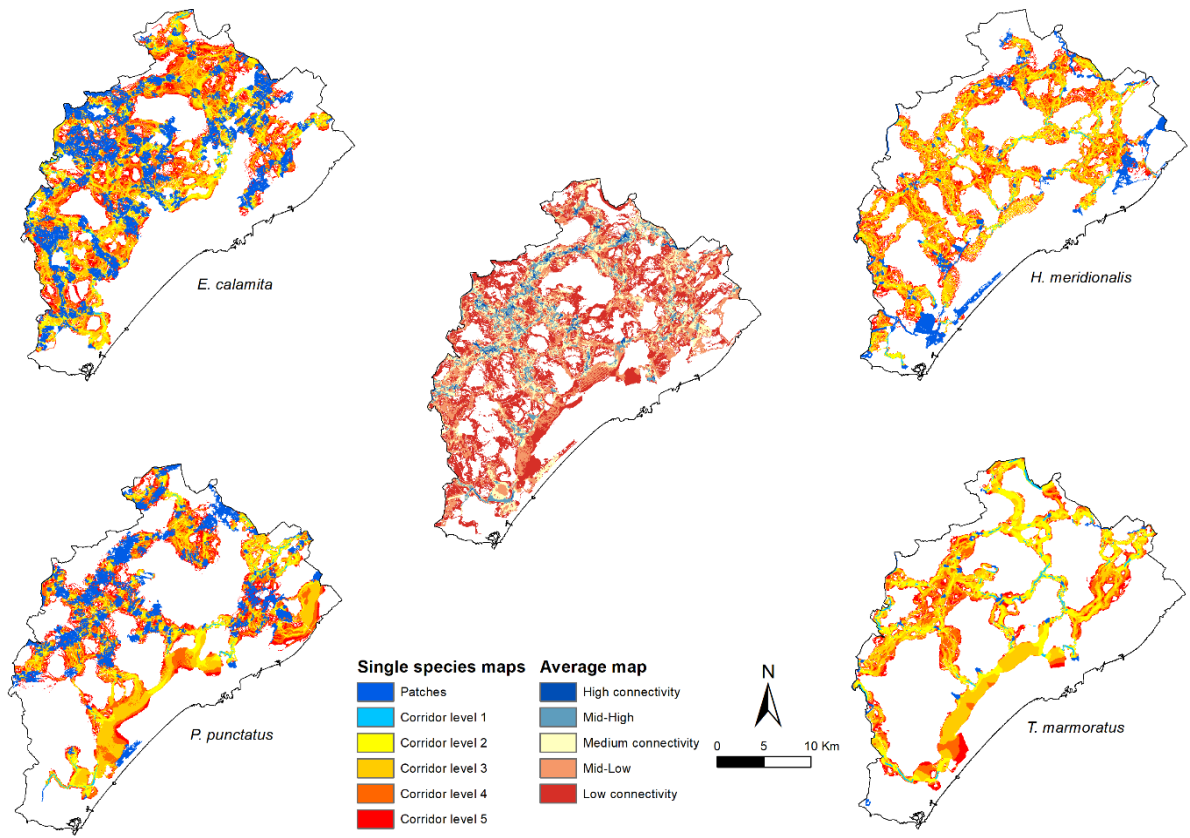

**Fig. 1** Single maps of connectivity levels and average map for low dispersal group

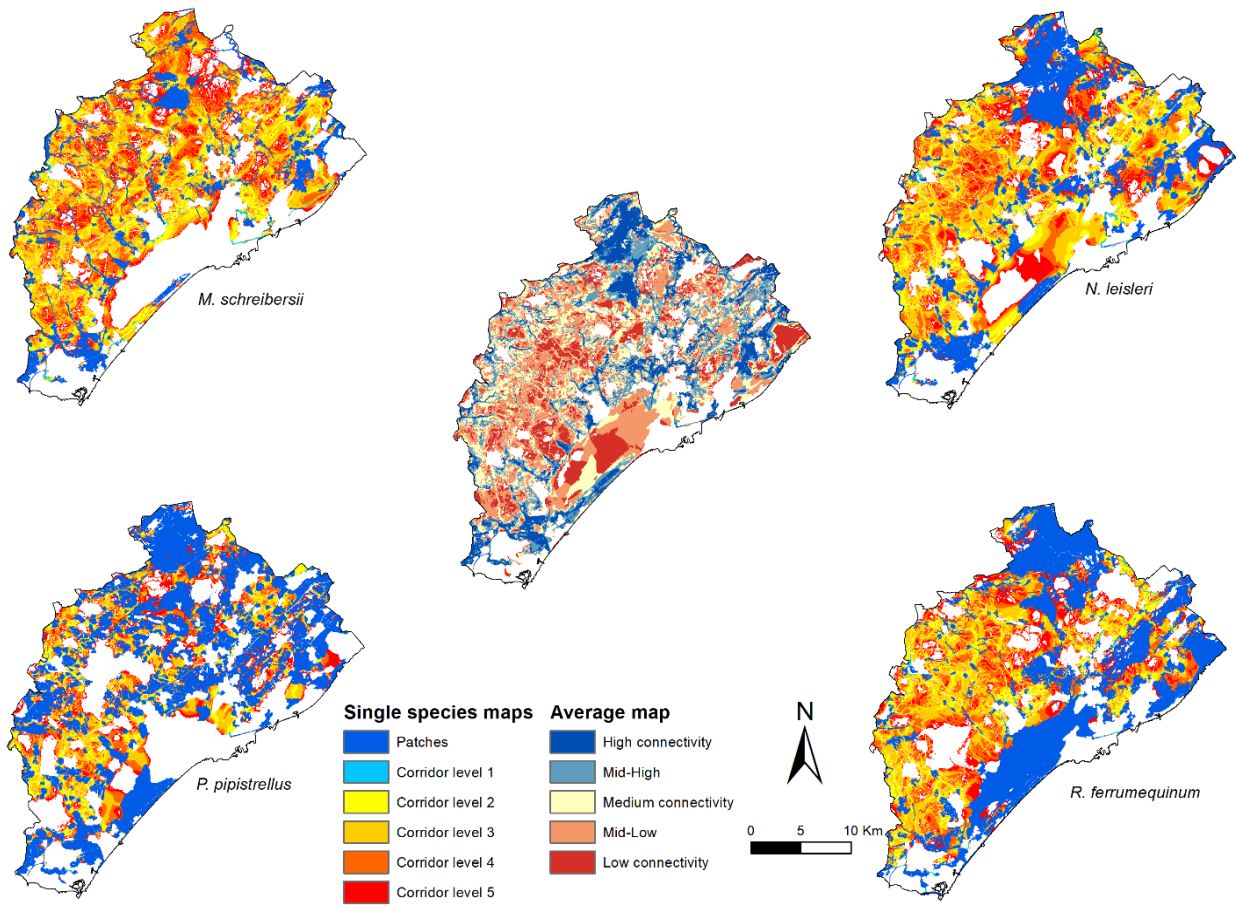

**Fig. 2** Single maps of connectivity levels and average map for high dispersal group

**Table 1** Percent of the study area covered by each connectivity levels per species

| Group | Species | Connectivity level 0 (% study area) | Connectivity level 1 (% study area) | Connectivity level 2 (% study area) | Connectivity level 3 (% study area) | Connectivity level 4 (% study area) | Connectivity level 5 (% study area) | Connectivity level 6 -7 (% study area) |
|---|---|---|---|---|---|---|---|---|
| | | | | | | | | |

| | | | | | | | |
|---|---|---|---|---|---|---|---|
| Low dispersal | *Triturus marmoratus* | 0,81 | 1,01 | 7,71 | 10,88 | 6,69 | 3,80 | 69,10 |
| | *Hyla meridionalis* | 5,15 | 0,78 | 6,75 | 11,18 | 9,03 | 5,58 | 61,53 |
| | *Epidalea calamita* | 15,89 | 0,64 | 6,50 | 11,72 | 10,56 | 7,01 | 47,69 |
| | *Pelodytes punctatus* | 10,55 | 0,47 | 4,98 | 10,03 | 8,68 | 6,53 | 58,76 |
| | | | | | | | | |
| High dispersal | *Rhinolophus ferrumequinum* | 30,85 | 0,53 | 7,04 | 15,21 | 13,51 | 7,85 | 25,00 |
| | *Miniopterus schreibersii* | 12,67 | 0,68 | 9,97 | 20,40 | 16,39 | 8,68 | 31,21 |
| | *Nyctalus leisleri* | 21,51 | 0,65 | 9,98 | 19,50 | 15,03 | 8,20 | 25,13 |

| | Pipistrellus pipistrellus | 36,75 | 0,41 | 4,72 | 9,08 | 8,45 | 6,22 | 34,38 |

**Table 2** Percent of the study area covered by each connectivity levels for average maps of species dispersal groups

| Group | High connectivity: level 0-1 (% of study area) | Level 1-2 (% of study area) | Level 2-3 (% of study area) | Level 3-4 (% of study area) | Low connectivity: level 4-5 (% of study area) | No connectivity: level 5-7 (% of study area) |
|---|---|---|---|---|---|---|
| Low dispersal species | 0,75 | 4,25 | 9,39 | 15,90 | 25,05 | 44,66 |
| High dispersal species | 12,56 | 14,06 | 20,51 | 21,10 | 13,60 | 18,17 |
| All species | 0,72 | 5,48 | 17,75 | 30,53 | 24,14 | 21,38 |

Online Resource 7: SD maps for the three groups

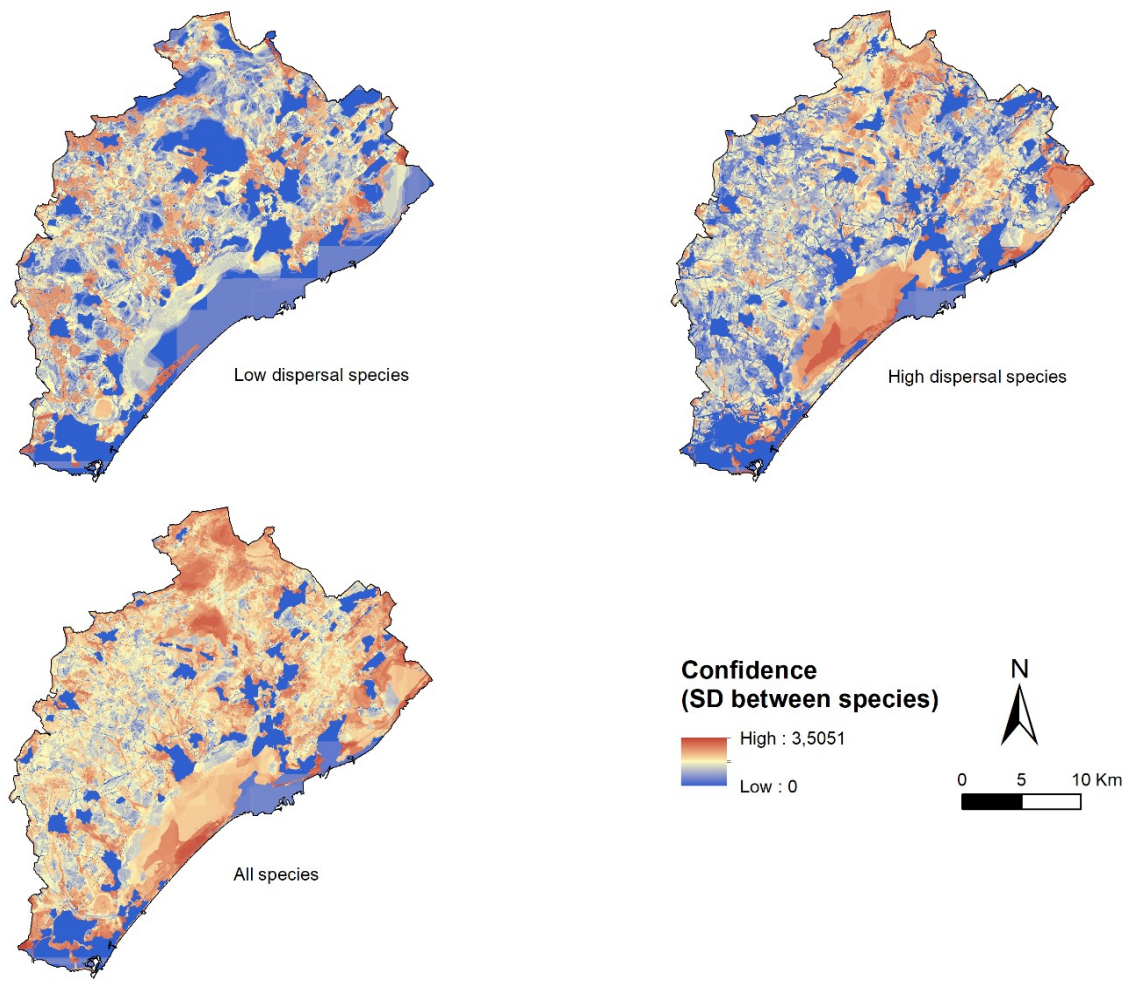

**Fig. 1** SD maps of the three dispersal groups (low dispersal species, high dispersal species, and all species

Online Resource 8: Connectivity within urban sprawl projects

**Table 1** Amount of surface area (in ha) covered by connectivity levels within urban sprawl projects for high dispersal species (high disp sp), low dispersal species (low disp sp) and all species (all sp), project numbers refer to figure 6

|  | Projects | 1 | 2 | 3 | 4 | 5 | 6 |
|---|---|---|---|---|---|---|---|
| **High connectivity** | High disp sp | 0.3081265 | 0 | 10.072808 | 0 | 0 | 0 |
|  | Low disp sp | 0 | 0 | 0 | 0 | 0 | 0 |
|  | All sp | 0 | 0 | 0 | 0 | 0 | 0 |
| **Mid-High connectivity** | High disp sp | 8.0005514 | 0.52 | 23.523383 | 0 | 0 | 0 |
|  | Low disp sp | 0 | 0 | 0 | 0 | 0 | 0 |
|  | All sp | 0 | 0 | 0 | 0 | 0 | 0 |
| **Medium connectivity** | High disp sp | 5.5019522 | 7.22 | 17.554427 | 0 | 0 | 0 |
|  | Low disp sp | 0 | 0 | 0 | 0 | 0 | 0 |
|  | All sp | 8.6334838 | 0 | 6.6726237 | 0 | 0 | 0 |
| **Mid-Low connectivity** | High disp sp | 2.2381891 | 7.82 | 1.9103252 | 0 | 0 | 0 |
|  | Low disp sp | 9.5138646 | 0 | 0.2391234 | 0 | 0 | 0 |
|  | All sp | 6.179324 | 3.66 | 43.317784 | 0 | 0 | 0 |
| **Low connectivity** | High disp sp | 3.8682172 | 5.92 | 0.7846778 | 0 | 0 | 0 |
|  | Low disp sp | 9.5977501 | 0 | 25.688742 | 0 | 0 | 0 |
|  | All sp | 4.7066528 | 12.7 | 3.6469772 | 0 | 0 | 0 |
| **No connectivity** | High disp sp | 30.822174 | 4.51 | 10.881222 | 1.6582379 | 0.42 | 1.7041568 |
|  | Low disp sp | 31.627596 | 26 | 38.798977 | 1.6582379 | 0.42 | 1.7041568 |
|  | All sp | 31.219751 | 9.63 | 11.089461 | 1.658239 | 0.42 | 1.7041573 |